\documentclass[conference,compsoc,a4paper]{IEEEtran}

\ifCLASSINFOpdf
\else
\fi

\usepackage{graphicx}
\usepackage{amsmath}
\usepackage{url}
\usepackage{array}

\newcommand{\old}[1]{{}}

\makeatother

\usepackage{algpseudocode}
\usepackage{algorithm}
\usepackage{smartdiagram}
\usepackage{cite}
\usepackage{pgfplotstable}
\pgfplotsset{compat=1.4}
\usepackage{soul}
\usepackage{multirow}
\graphicspath{{Figures/}}
\usepackage{subcaption}
\usepackage{breakurl}
\usepackage[breaklinks]{hyperref}

\usepackage{pifont}
\usepackage{amssymb}
\newcommand{\cmark}{\ding{51}}%
\newcommand{\xmark}{\ding{55}}%
\newcommand{\mycomment}[1]{}

\begin{document}

\title{When a RF Beats a CNN and GRU, Together - A Comparison of Deep Learning and Classical Machine Learning Approaches for Encrypted Malware Traffic Classification}
\author{
		\IEEEauthorblockN
		{
			Adi Lichy\IEEEauthorrefmark{1}\IEEEauthorrefmark{3},
			Ofek Bader\IEEEauthorrefmark{1}\IEEEauthorrefmark{3},
			Ran Dubin\IEEEauthorrefmark{1}\IEEEauthorrefmark{3},
			Amit Dvir\IEEEauthorrefmark{1}\IEEEauthorrefmark{3},
			Chen Hajaj\IEEEauthorrefmark{2}\IEEEauthorrefmark{3}\IEEEauthorrefmark{4}\\
		}
		\IEEEauthorblockA{\IEEEauthorrefmark{3}Ariel Cyber Innovation Center, Ariel University, Israel}
		\IEEEauthorblockA{\IEEEauthorrefmark{1}Department of Computer Science, Ariel University, Israel}
	    \IEEEauthorblockA{\IEEEauthorblockA{\IEEEauthorrefmark{4}Department of Industrial Engineering \& Management, Ariel University, Israel}
	    \IEEEauthorblockA{\IEEEauthorrefmark{2}Data Science and Artificial Intelligence Research Center, Ariel University, Israel}
	    }

} 
\maketitle  
\thispagestyle{plain}
\pagestyle{plain}
\begin{abstract}
Internet traffic classification is widely used to facilitate network management. It plays a crucial role in Quality of Services (QoS), Quality of Experience (QoE), network visibility, intrusion detection, and traffic-trend analyses. While there is no theoretical guarantee that deep learning (DL)-based solutions perform better than classic machine learning (ML)-based ones, DL-based models have become the common default.  
This paper compares well-known DL-based and ML-based models and shows that in the case of malicious traffic classification, state-of-the-art DL-based solutions do not necessarily outperform the classical ML-based ones. We exemplify this finding using two well-known datasets for a varied set of tasks, such as: malware detection, malware family classification, detection of zero-day attacks, and classification of an iteratively growing dataset. Note that, it is not feasible to evaluate all possible models to make a concrete statement, thus, the above finding is not a recommendation to avoid DL-based models, but rather empirical proof that in some cases, there are more simplistic solutions, that may perform even better. 

\end{abstract}


\section{Introduction}
Internet traffic classification has been widely used to facilitate network management. It plays a crucial role in Quality of Services (QoS), Quality of Experience (QoE), network visibility, intrusion detection, and traffic trend analyses. To improve privacy, integrity, confidentiality, and protocol obfuscation, today’s traffic is based on encryption protocols, e.g., SSL/TLS \cite{cite1rfc8446}. While encryption is beneficial from the user's perspective, many entities find the ongoing transition to more privacy-preserving protocols less preferable. For example, network providers are interested in various QoE aspects, such as video resolution and stalling events (rebuffering) of a user, which are encrypted and, therefore, concealed from them.

From a security point of view, encryption protocols make it harder to classify malicious traffic or exfiltration traffic that endangers users. Therefore, efficient classification of encrypted traffic has become a topic for discussion that has attracted many researchers in the fields of QoS, QoE, and cybersecurity \cite{cite3, cite6, cite10}. 

Most available studies tackle the classification problem using different approaches such as port-based, payload-based and behavior-based. Moreover, they categorize data representation methods in various ways. For instance, behavior-based representation aims to capture the patterns of interaction between different network elements. In contrast, network flows are represented as vectors of statistical features. 

Previous works have shown that classical machine learning models are also applicable in the scope of  QoE \cite{8486321}, mobile app fingerprinting \cite{flowprint}; or try to classify user activities \cite{M_actions, cite24, cite25} and webpage fingerprinting \cite{cite15}. 
Recently, several works transformed the flow into a language to use word embedding and other Natural Language Processing (NLP) techniques\cite{DVIR_cluster_unknonw}. In contrast, others converted the flow to a traffic image to harness image processing techniques and equivalent Deep Learning (DL) architectures \cite{ cite3, cite6, cite10, FlowPic2021}. A survey about the encrypted traffic techniques and methods can be found in \cite{papadogiannaki2021survey}

The same follows from the perspective of cybersecurity, where 
evasive attackers take advantage of encryption. Attackers can bypass most inspection devices to deliver malware inside the encrypted network and use the encrypted data to exfiltrate precious data outside the organization.
Threat research \cite{f5networks} concluded that 71\% of malware uses encryption to hide when it communicates back to command and control locations. 
Cybercriminals take advantage of this to hide their communication from network analysis devices. 
Therefore, a vast amount of previous work on malware traffic classification focused on the problem of binary classification of network traffic (benign or malicious), while classifying malware traffic into malware families \cite{cite26, cite27, cite29, cite30,maldist_ccnc}. 

In this paper, we compare well-known ML and DL-based models for classification of malicious traffic. Thus, the main research in this work focuses on "\textit{whether a complex DL architecture outperforms a classic ML model for this classification task, or perhaps a simple ML model is more than sufficient}".

The remainder of this paper is organized as follows. First, in Section \ref{sec:RW} the related work is reviewed and discussed. Second, our datasets sources, prepossessing and experimental design are presented in Section \ref{sec:ED}. The results of our experiments are discussed and evaluated in Sections \ref{sec:ER}. Finally, we conclude with a discussion and directions for future research in Section \ref{sec:con}.


\section{Related Work}
\label{sec:RW}
Classification of encrypted traffic is a critical issue today. From past to present, researchers have performed
various studies on this subject. A recent survey on ML for networking includes traffic prediction, routing and classification, congestion control, resource and fault management, QoS and QoE management, and network security is presented in \cite{cite3deepmal}. Other works on network measurement problems such as network anomaly detection can be found in \cite{cite5deepmal, cite6deepmal} - including ML-based approaches \cite{cite4deepmal}, ML for network traffic classification \cite{cite8deepmal} and network security \cite{cite7deepmal, Wang_Mal_2018}.

Recently, DL approaches have begun to be used with promising performance results, mainly associated with traffic classification tasks \cite{Wang1DCNN, DeepMAL, FlowPic2021, LopezNetworkTrafficClassifier, Distiller,GrayPicFingerprintin, cite21deepmal, cite16seq2img} and malware traffic classification \cite{WeiWangMalwareTrafficClassification, NetML, UnknownMalwareDetectionBenGurion, EncryptedMalwareContext, MTAKDD19, DeepMAL, yesML, cite27, cite29, cite30, citeMTATLS}. Some of the works extracted features from early parts of a session, such as the first payload bytes of the session \cite{WeiWangMalwareTrafficClassification, Wang1DCNN}, while others used the first packets of the network session \cite{DeepMAL, LopezNetworkTrafficClassifier}. 
Other works, such as \cite{FlowPic2021} and \cite{GrayPicFingerprintin} used part or a whole session to create a histogram represented as an image, which is later fed into a Convolutional Neural Network (CNN).

Lotfollahi \textit{et al.} \cite{cite12deepmal} used SAE and 1D-CNN at the packet level. In contrast, Lopez-Martin \textit{et al.} \cite{LopezNetworkTrafficClassifier} presented different DL architectures based on CNN and LSTM networks to perform encrypted traffic classification. 


Wang \textit{et al.} \cite{Wang1DCNN} used DL based on raw traffic data for the task of normal traffic detection, where the general idea is that the need for feature extraction is obsolete, and the classifier can work as is by viewing the network stream. To test this, they performed multiple experiments while keeping the shape of the $784$ bytes as a single vector, to represent sequential data, and the DL architecture consists of 1D-Convolutional Layers instead of 2D, it denoted \textit{M1CNN}.

Shapira \& Shavitt \cite{FlowPic2021} proposed a generic representation of network traffic names termed FlowPic. FlowPic is a 2-dimensional histogram of packet sizes over time, where a pixel/cell at index ($s$, $t$) in the image, counts the number of packets with size $s$ at relative time $t$.
The histograms do not contain the packets of the whole session. Instead, they contain only packets that fall inside a predefined time-window.

Wang \textit{et al.} \cite{WeiWangMalwareTrafficClassification}
move away from the traditional careful handy-crafted features of classical ML towards representation learning of raw traffic data with DL in the task of malware traffic detection. They proposed extracting the first $784$ payload bytes of a session as raw data and then reshaping them into a [$28X28$] image to feed it into a customized Deep Learning architecture of the known CNN LeNet-5 architecture. As the proposed DL architecture consists of 2D-Convolutional Layers, it denoted \textit{M2CNN}. 

The features \cite{WeiWangMalwareTrafficClassification} were later adopted by Aceto \textit{et al.} \cite{Distiller} to create a multi-modal multi-task Deep Learning architecture for the problem of classifying network sessions in multiple levels (or tasks). \textit{DISTILLER} is a multi-modal multi-task Deep Learning architecture. The architecture aims to solve multiple encrypted traffic classification tasks simultaneously with different multiple data modalities as input. In the pre-training phase, each modality trains to solve all tasks and learns intra-modality relations and dependencies. In contrast, in the fine-tuning phase, all of the modalities are combined to form the \textit{DISTILLER} and learn inter-modality relations and dependencies.

Marin \textit{et al.} \cite{DeepMAL} describes two variants of malware classification, a packet-based one and a session-based one. The session-based variant architecture leverages a convolutional layer and fully connected layers to predict and classify malware. The extracted features from each session are $n$ payload bytes (per packet) from the first $m$ packets. In our paper we used only the session-based variant architecture, it denoted \textit{DeepMal}.

Yang \textit{et al.} \cite{DeepLearning-Zero-Day} showed that CNN had higher results than an XGBoost for zero-day on the applications. In our case with XGBoost has far lower result than RF.

Bader \textit{et al.} \cite{maldist_ccnc}, presented a novel Deep Learning network architecture (MalDIST) and showed that the novel approach outperformed the other state-of-the-art DL models in the case of malware network classification.

\section{Experimental Design}
\label{sec:ED}
\subsection{Datasets}
For our experiments, we created three datasets based on a combination of the following five publicly available datasets (as described in Table~\ref{table:our-datasets}): 
\begin{itemize}
    \item \textbf{StratosphereIPS}\cite{stratodatasets} is a dataset generated by Stratosphere, which was used for samples of benign traffic. For works that adopted this dataset see:~\cite{Ransomware_pre-encrypted_alert,Decetion_ENC_MAL_ML,NetML}.
    \item \textbf{ISCX2016 (VPN-nonVPN2016)}\cite{ISCX2016, ISCX2016_Dataset} is a dataset that consists of different types of benign traffic and applications. The samples are labeled according to the application (e.g., Facebook, YouTube, Spotify, etc.) and traffic type category (e.g., streaming, VoIP, chat, etc.) along with encapsulation label (VPN/non-VPN).  For  works that adopted this dataset see:~\cite{FlowPic2021,NetML,Distiller}.
    \item \textbf{Ariel (BOA)}\cite{BOA_conf} is a benign dataset that was collected over a period of more than two months. The dataset contains label tuples of the operating system, browser, and applications such as YouTube and Facebook. See: \cite{ShahbazHighClassification,Yang2020IEEE} for works that adapted this dataset.
    \item \textbf{Malware-Traffic-Analysis.net (MTA)}\cite{MTA} is a website that shares many types of malware infection traffic (e.g., ransomware and exploit kits) for analysis. Since 2013, the dataset has been updated daily with relevant up-to-date malware traffic. Specifically, our collected malware are from August 2021. Every binary file in the PCAPs has been confirmed as malicious by Intrusion-Detection Systems (IDS) and Antivirus software. Papers such as \cite{citeMTATLS, MTAKDD19} used this dataset for malware detection tasks.
    \item \textbf{USTC-TFC2016 (USTC)}\cite{USTC} contain malware and benign samples from the University of Science and Technology of China (USTC). The benign samples that were used are applications such as Webio, World of Warcraft, Facetime, and Wechat. As for the malware, the samples belong to various families such as Miuref, Neris, Nsis, and Shifu. This Dataset was used by~\cite{WeiWangMalwareTrafficClassification,Wang1DCNN,hwang2020unsupervised}.
\end{itemize}
The five datasets were used to evaluate both classic ML and DL-based models for malware detection and family classification tasks as explained in the following section.
Naturally, one would expect that the malicious dataset would be composed of both MTA and USTC datasets. Still, our initial analysis showed that both datasets include mutual malware families (e.g., Dridex and Cridex, Emotet and Geodo). Thus, before combining the datasets, we verified that learning samples of malware families from one dataset would allow the classifier to correctly classify the family's instances on the other dataset. 
The results of this analysis are reported in Table~\ref{table:MTAvsUSTC}. Note that  0\% accuracy was the result, for all models, on samples of both families when the training set was composed of samples from USTC and the test set was composed of samples of the same family on MTA. Deep inspection of the samples showed another interesting phenomenon. The level of using TLS in the malware traffic is different between the two datasets (see Table \ref{table:TLS}). Thus, to provide an accurate analysis, we analyzed each malicious dataset independently and combined the two datasets, where the two similar malware families were considered different malware. For the benign traffic, all in all, we used benign samples from four datasets: StratosphereIPS\footnote{The PCAPs that we selected from StratosphereIPS were the same as in NetML~\cite{NetML}.}, ISCX2016, BOA, and USTC-TFC2016.


\begin{figure*}[t!]
\centering
\subfloat[MTAB - Before filtering\label{fig:mtab-dataset-dist-precleaning}]{%
      \includegraphics[width= 4cm,height=4cm]{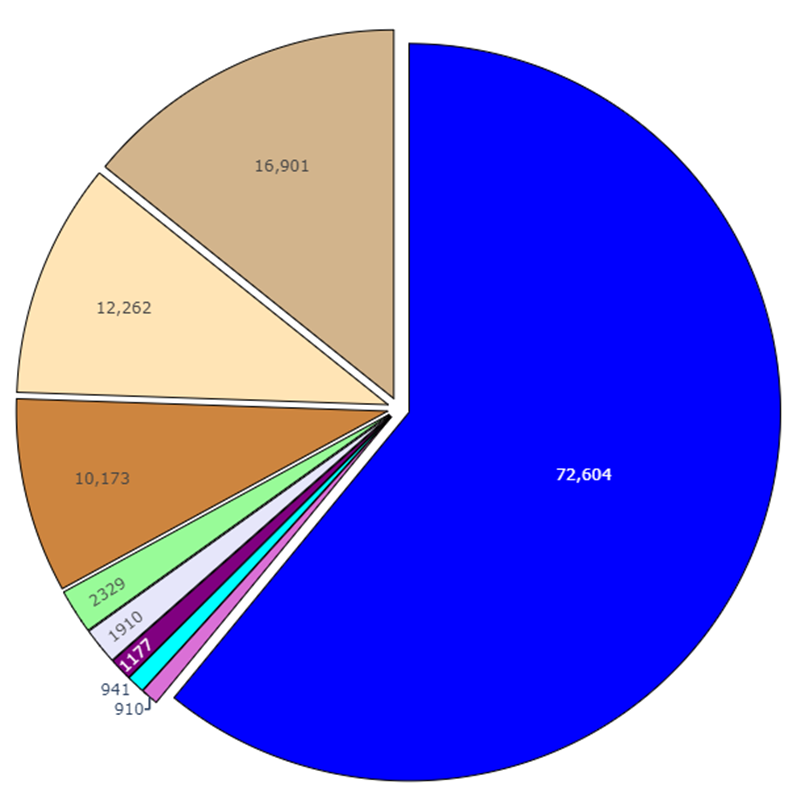}
    }
    \hfill
\subfloat[MTAB - After filtering\label{fig:mtab-dataset-dist-postcleaning}]{%
      \includegraphics[width=4cm,height=4cm]{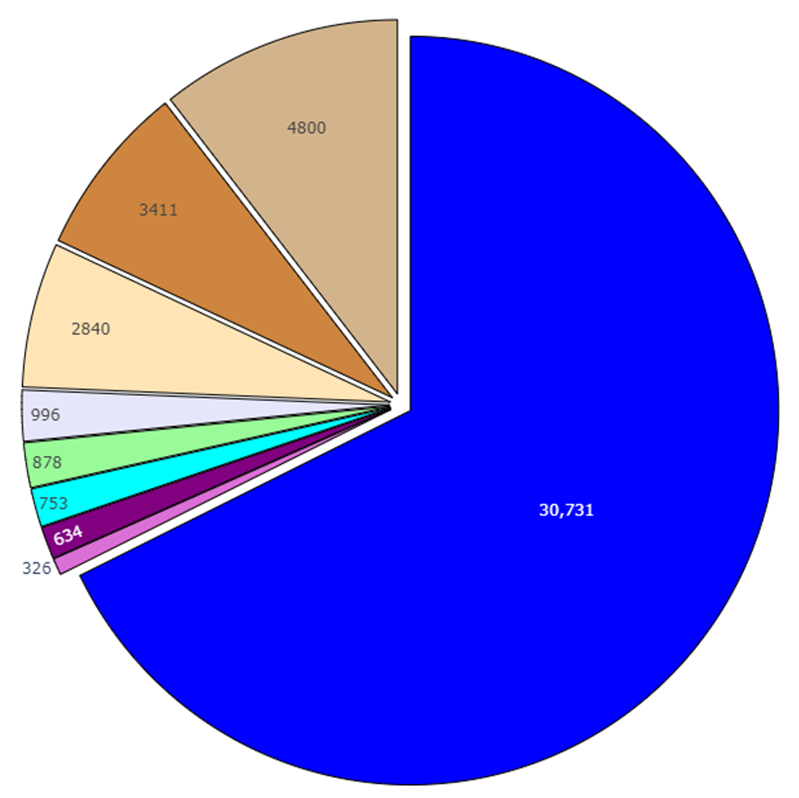}
    }
    \hfill
\subfloat[MTAB - After  balancing\label{fig:mtab-dataset-dist-balance}]{%
      \includegraphics[width=4cm,height=4cm]{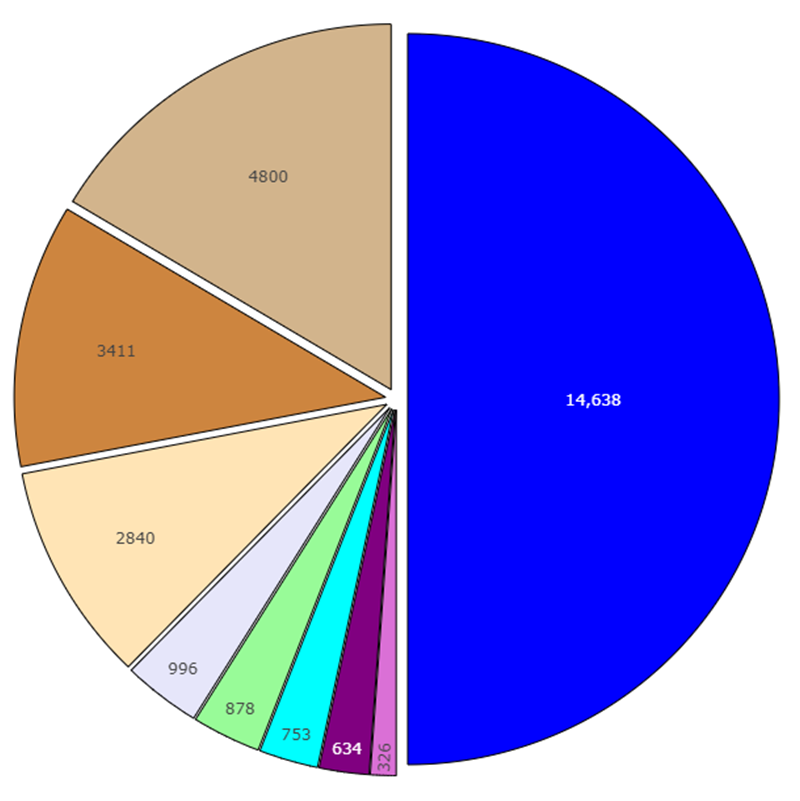}
    }

\subfloat[USTCB - Before filtering\label{fig:ustcb-dataset-dist-precleaning}]{%
      \includegraphics[width=4cm,height=4cm]{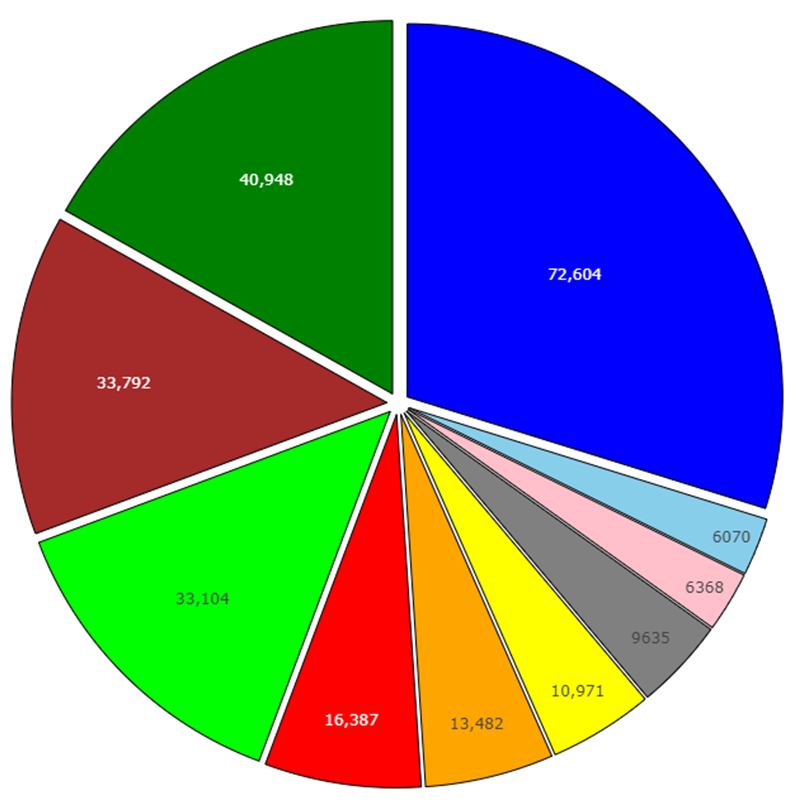}
    }
        \hfill
\subfloat[USTCB - After filtering\label{fig:ustcb-dataset-dist-postcleaning}]{%
      \includegraphics[width=4cm,height=4cm]{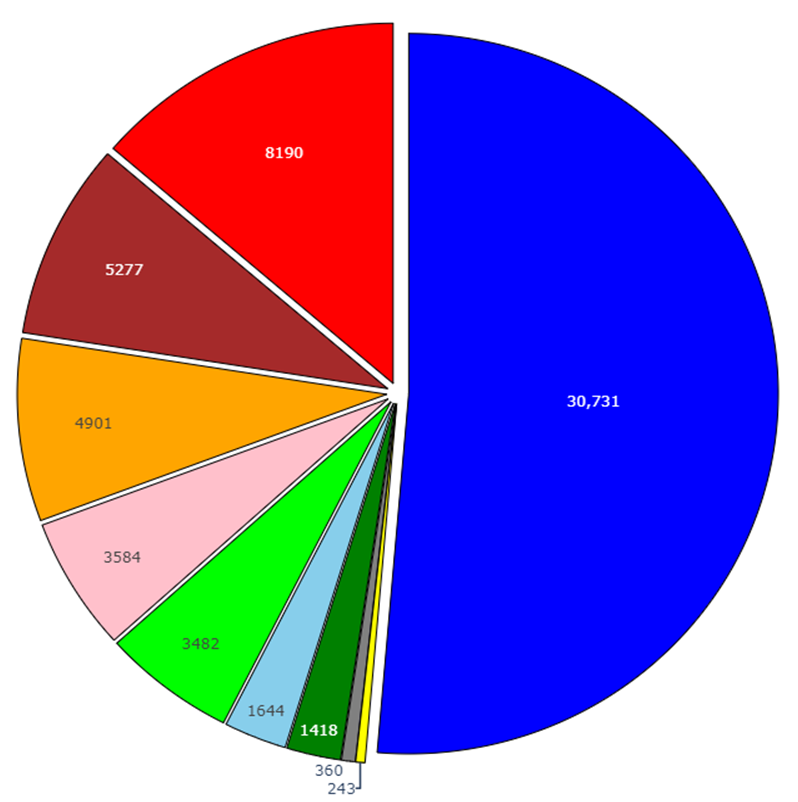}
    }
   \hfill
\subfloat[USTCB - After balancing\label{fig:ustcb-dataset-dist-balance}]{%
      \includegraphics[width=4cm,height=4cm]{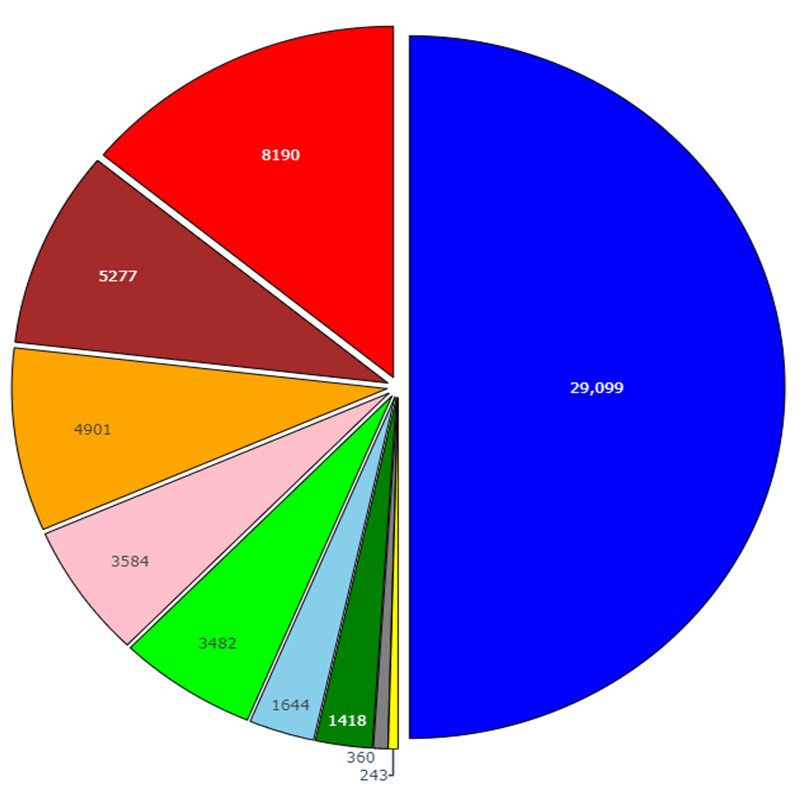}
    }

\subfloat[MUB - Before filtering\label{fig:mub-dataset-dist-precleaning}]{%
      \includegraphics[width=4cm,height=4cm]{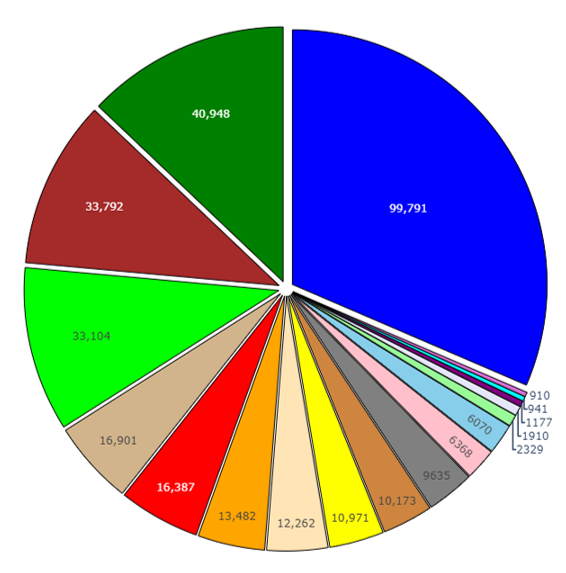}
    }
     \hfill
\subfloat[MUB - After filtering\label{fig:mub-dataset-dist-postcleaning}]{%
      \includegraphics[width=4cm,height=4cm]{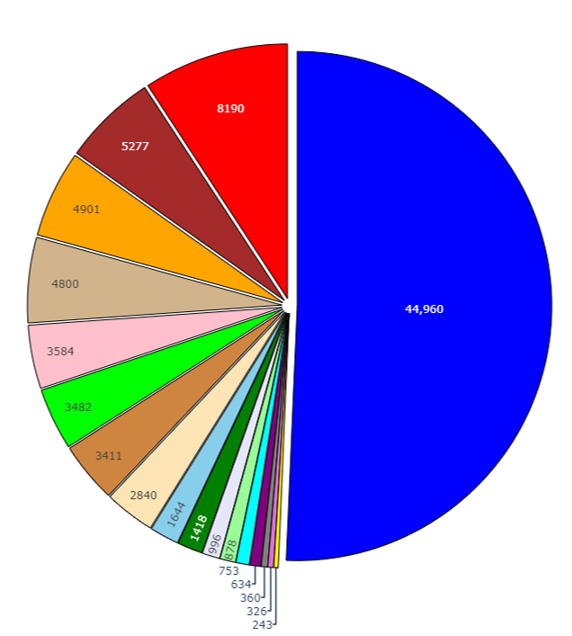}
    }
 \hfill
\subfloat[MUB - After balancing\label{fig:mub-dataset-dist-balance}]{%
      \includegraphics[width=4cm,height=4cm]{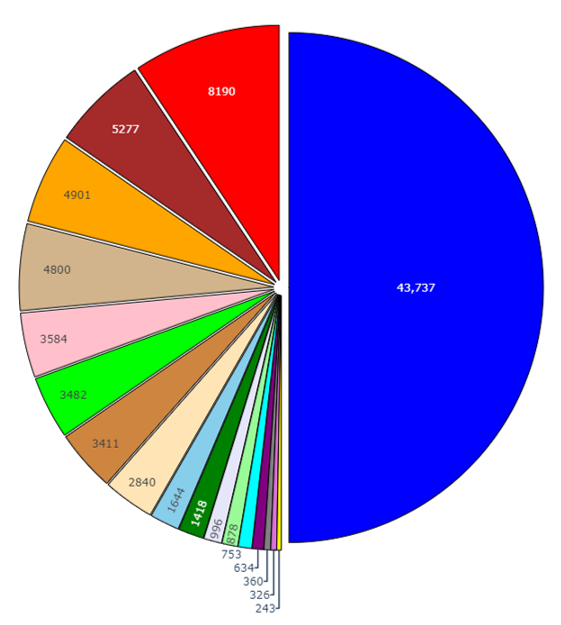}
    }
\label{fig:dataset-dist}
\centering
\includegraphics[width=0.5\textwidth]{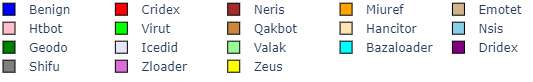}
\caption{Number of sessions in our datasets (per class): before filtering, after filtering, and after balancing.}
\end{figure*}

\begin{table}[H]
\caption{Datasets correlations, training on one dataset while testing with the other}.
\label{table:MTAvsUSTC}
\begin{center}
\begin{tabular}{ | c | c | c | c | c |}
\hline
Malware Family & Source & Test &  Model & Accuracy \\
\hline
\multirow{8}{*}{Dridex/Cridex} & \multirow{4}{*}{MTA} & \multirow{4}{*}{USTC} & Random Forest & 86.69\% \\ \cline{4-5}
& & & LGBM & 62.15\% \\ \cline{4-5}
& & & XGboost & 48.36\%  \\ \cline{4-5}
& & & Extra Trees & 99.05\%  \\ \cline{2-5}
& \multirow{4}{*}{USTC} & \multirow{4}{*}{MTA} & Random Forest & 00.00\% \\ \cline{4-5}
& & & LGBM & 00.00\% \\ \cline{4-5}
& & & XGboost & 00.00\% \\ \cline{4-5}
& & & Extra Trees & 00.00\% \\
\hline
\multirow{8}{*}{Emotet/Geodo} & \multirow{4}{*}{MTA} & \multirow{4}{*}{USTC} & Random Forest & 01.85\% \\ \cline{4-5}
& & & LGBM & 05.13\% \\ \cline{4-5}
& & & XGboost & 06.42\%  \\ \cline{4-5}
& & & Extra Trees & 00.13\%  \\ \cline{2-5}
& \multirow{4}{*}{USTC} & \multirow{4}{*}{MTA} & Random Forest & 00.49\% \\ \cline{4-5}
& & & LGBM & 01.42\% \\ \cline{4-5}
& & & XGboost & 00.42\% \\ \cline{4-5}
& & & Extra Trees & 00.44\% \\
\hline
\end{tabular}
\end{center}
\end{table}

We created three datasets named MTAB, USTCB, and MUB (see Table~\ref{table:our-datasets} for the composition of each dataset). The first dataset, MTAB, contained only the MTA malware samples with benign samples from ISCX2016, StratoshpereIPS, and BOA. The second dataset, USTCB, contained only the USTC malware samples with benign samples from ISCX2016, StratoshpereIPS, and BOA. The third dataset, MUB, contained both MTA and USTC malware samples with benign samples from ISCX2016, StratoshpereIPS, BOA, and USTC.
In terms of malware families, we only chose families that contain at least 100 sessions (traffic) from MTA~\cite{MTA}, which resulted in 8 families. From the USTC dataset, which contained 10 malware families, we removed only one family, namely Tinba, because it contained only DNS traffic which was not part of the network traffic we aimed to classify (filtered by port 53).


\begin{table}[H]
\caption{Datasets - TLS traffic percentage}
\label{table:TLS}
\begin{center}
\begin{tabular}{ | c | c | c | c |}
\hline
Dataset & Malware/Benign & Normal & Encrypted\\
\hline
\multirow{2}{*}{MTAB} & Benign & 27.45\% & 72.55\% \\ \cline{2-4}
& Malware & 32.34\% & 67.66\% \\ 
\hline
\multirow{2}{*}{USTCB} & Benign & 13.81\% & 86.19\% \\ \cline{2-4}
& Malware & 95.47\% & 4.53\% \\ 
\hline
\multirow{2}{*}{MUB} & Benign & 25.9\% & 74.1\% \\ \cline{2-4}
& Malware & 74.34\% & 25.66\% \\ 
\hline
\end{tabular}
\end{center}
\end{table}

\begin{table}[H]
\caption{Our new datasets for evaluation}
\label{table:our-datasets}
\begin{center}
\begin{tabular}{ | m{1cm} | c | c  | c | c |}
\hline
  & MTA & USTC-M & ISCX + IPS + BOA & USTC-B \\
\hline
MTAB & \textcolor{green}{\cmark} & \textcolor{red}{\xmark} & \textcolor{green}{\cmark} &  \textcolor{red}{\xmark} \\
\hline
USTCB & \textcolor{red}{\xmark} & \textcolor{green}{\cmark} & \textcolor{green}{\cmark} & \textcolor{red}{\xmark} \\
\hline
MUB & \textcolor{green}{\cmark} & \textcolor{green}{\cmark} & \textcolor{green}{\cmark} & \textcolor{green}{\cmark} \\
\hline
\end{tabular}
\end{center}
\end{table}

\subsection{preprocessing}
Our preprocessing stage is comprised of three steps. First, we filtered sessions with less than 784 payload bytes, because they were not sufficiently informative enough~\cite{DeepMAL}, while~\cite{WeiWangMalwareTrafficClassification,Wang1DCNN} not filtered sessions but zero-padded them. This refinement step removed several sessions, as depicted in Figures \ref{fig:mtab-dataset-dist-postcleaning}, \ref{fig:ustcb-dataset-dist-postcleaning}, and \ref{fig:mub-dataset-dist-postcleaning}. Then, similar to other works \cite{Distiller, FlowPic2021}, we cleaned the datasets by removing sessions with irrelevant protocols such as SNMP, LLMNR and sessions that are considered noise, such as UDP broadcasts (e.g., Dropbox LAN Discovery). Finally, due to the imbalance between the number of benign and malware samples in the dataset, we randomly sampled the benign data such that the number of samples of benign and malware would be 50\% each, as illustrated in Figures~\ref{fig:mtab-dataset-dist-balance},~\ref{fig:ustcb-dataset-dist-balance}, and~\ref{fig:mub-dataset-dist-balance}.
As a result, the number of samples in the dataset was 29k for the MTAB dataset (Benign: 14638, Emotet: 4800, Qakbot: 3411, Hancitor: 2840, Icedid: 996...), 58k for the USTCB dataset (Benign: 29099, Cridex: 8190, Neris: 5277, Miuref: 4901, Htbot: 3584...) and 87k for the unified MUB dataset. Note that we did not balance the classes across malware families as the number of sessions varied considerably.

\begin{figure*}[!]
\centering
 \subfloat[MTAB
  \label{fig:MTAB-malware-family}]{%
\includegraphics[width=0.75\textwidth]{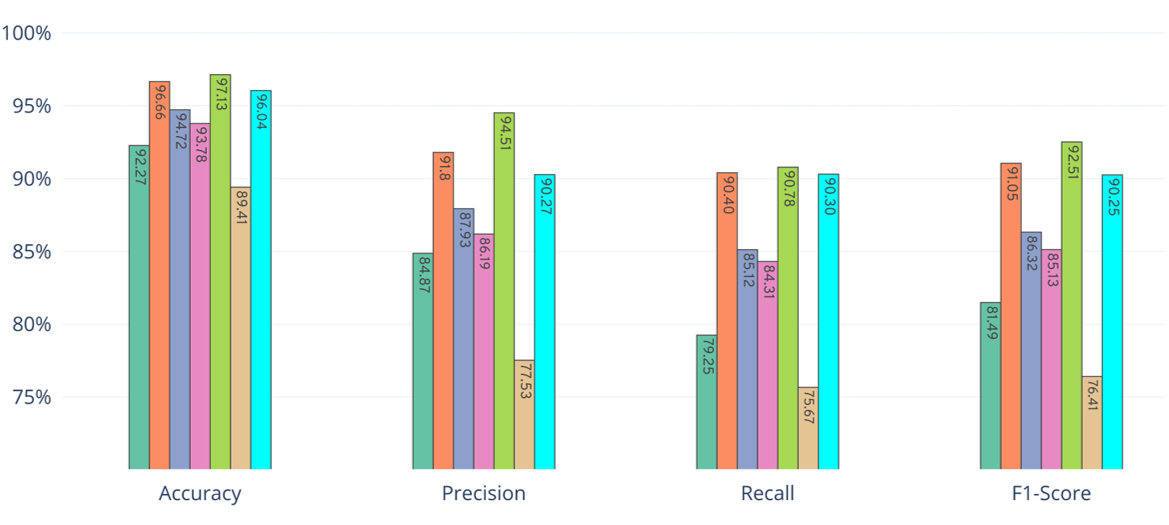}}
\hfill
\subfloat[USTCB\label{fig:USTCB-malware-family}]{%
  \includegraphics[width=0.75\textwidth]{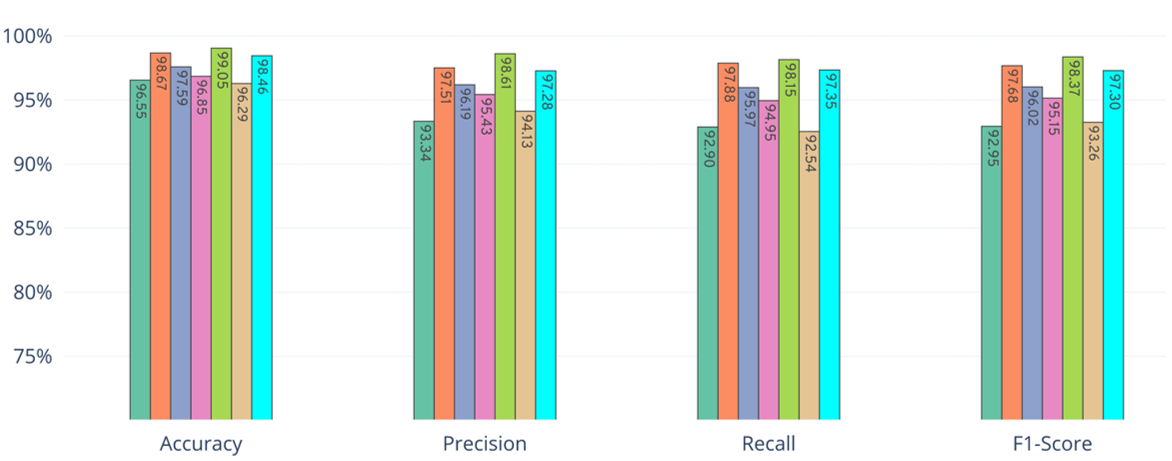}}
\hfill
\subfloat[MUB
  \label{fig:MUB-malware-family}]{%
\includegraphics[width=0.75\textwidth]{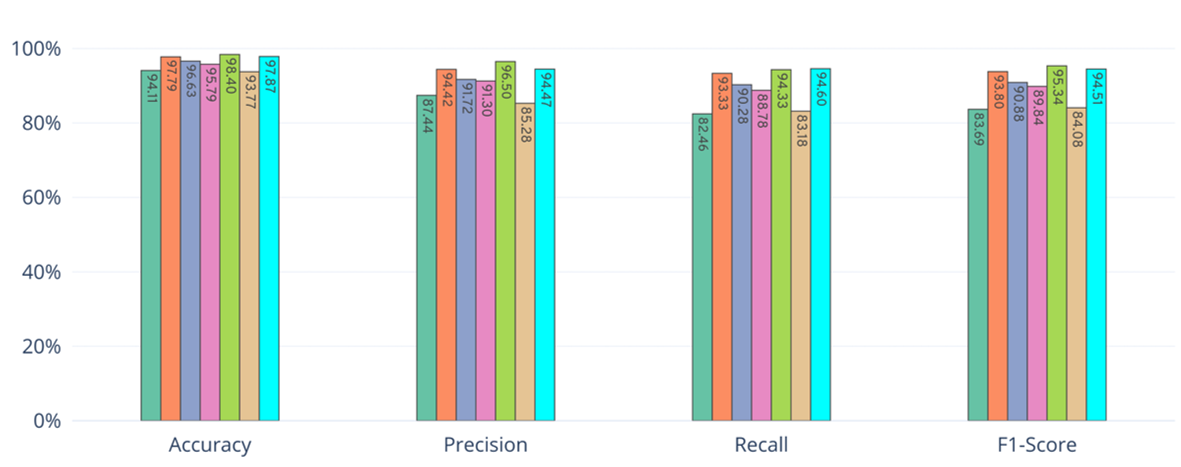}}
\label{fig:scores-family}

\centering
\includegraphics[width=0.4\textwidth]{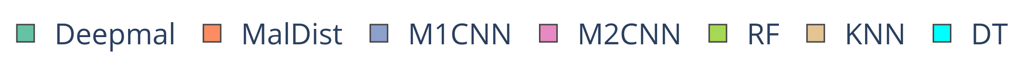}
\caption{Multi class malware traffic classification of the following models: \textit{DeepMAL} (green), \textit{MalDIST} (orange), \textit{M1CNN} (blue), \textit{M2CNN} (pink) \textit{RF} (light green), \textit{KNN} (brown) and \textit{DT} (cyan)}
\label{fig:malware-family-scores}
\end{figure*}

\begin{figure*}[!]
\centering
\hfill
\subfloat[\textit{MTAB MalDIST}.\label{fig:cf-matrices-mtab-maldist}]{%
  \includegraphics[width=.49\textwidth]{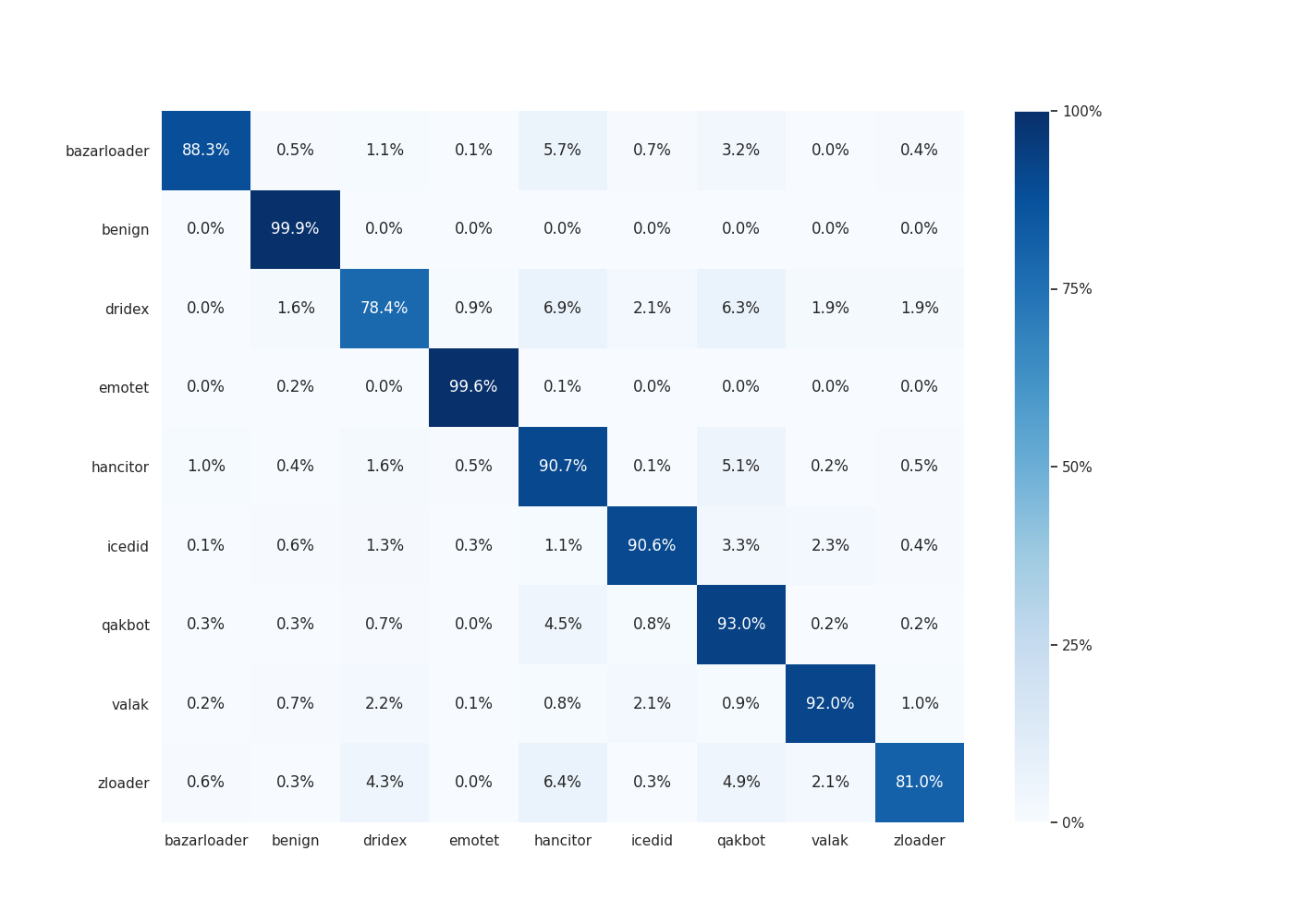}
}
\hfill
\subfloat[\textit{MTAB Random Forest}.\label{fig:cf-matrices-mtab-rf}]{%
  \includegraphics[width=.49\textwidth]{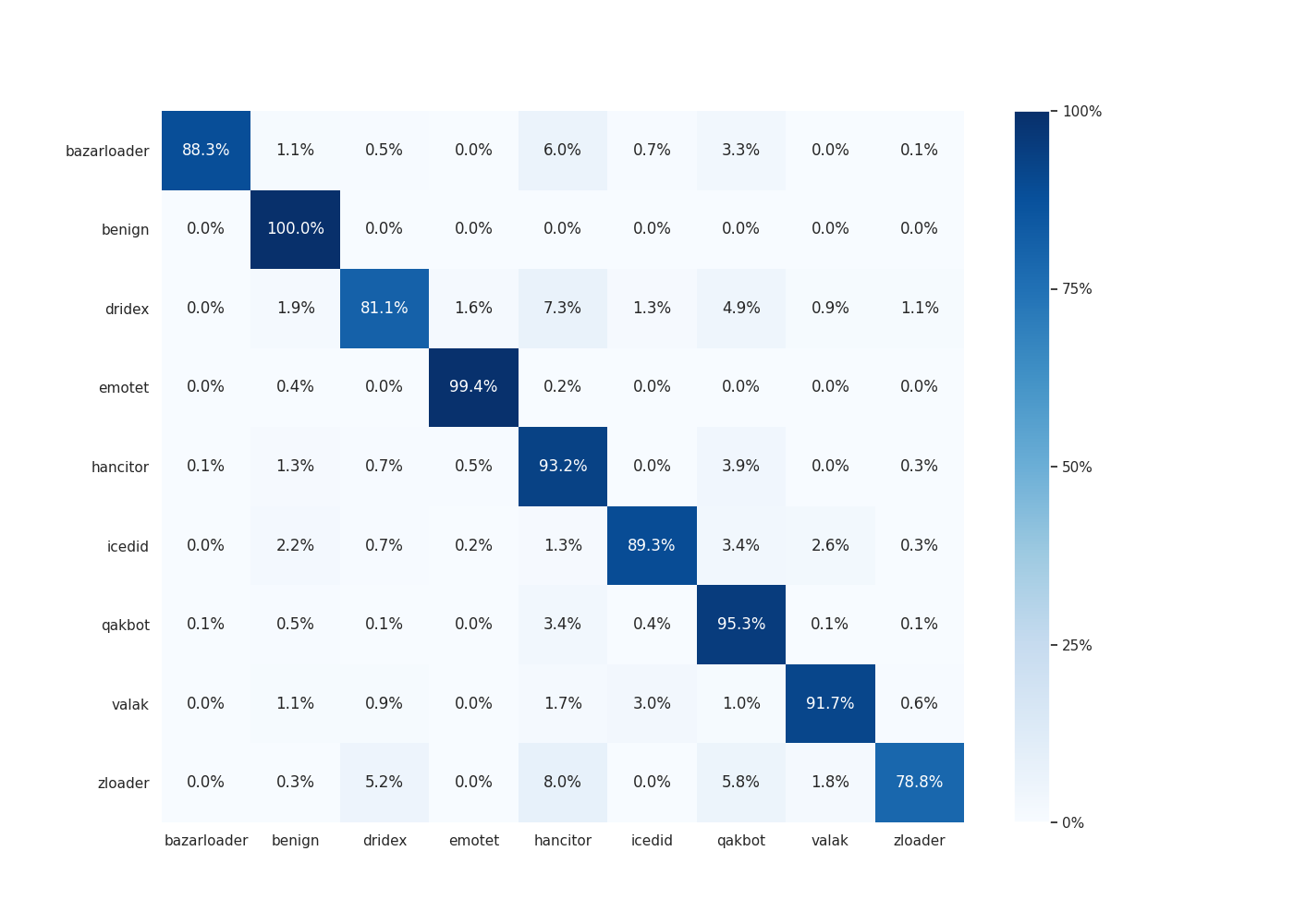}
}
\hfill
\subfloat[\textit{USTCB MalDIST}.\label{fig:cf-matrices-ustcb-maldist}]{%
  \includegraphics[width=.49\textwidth]{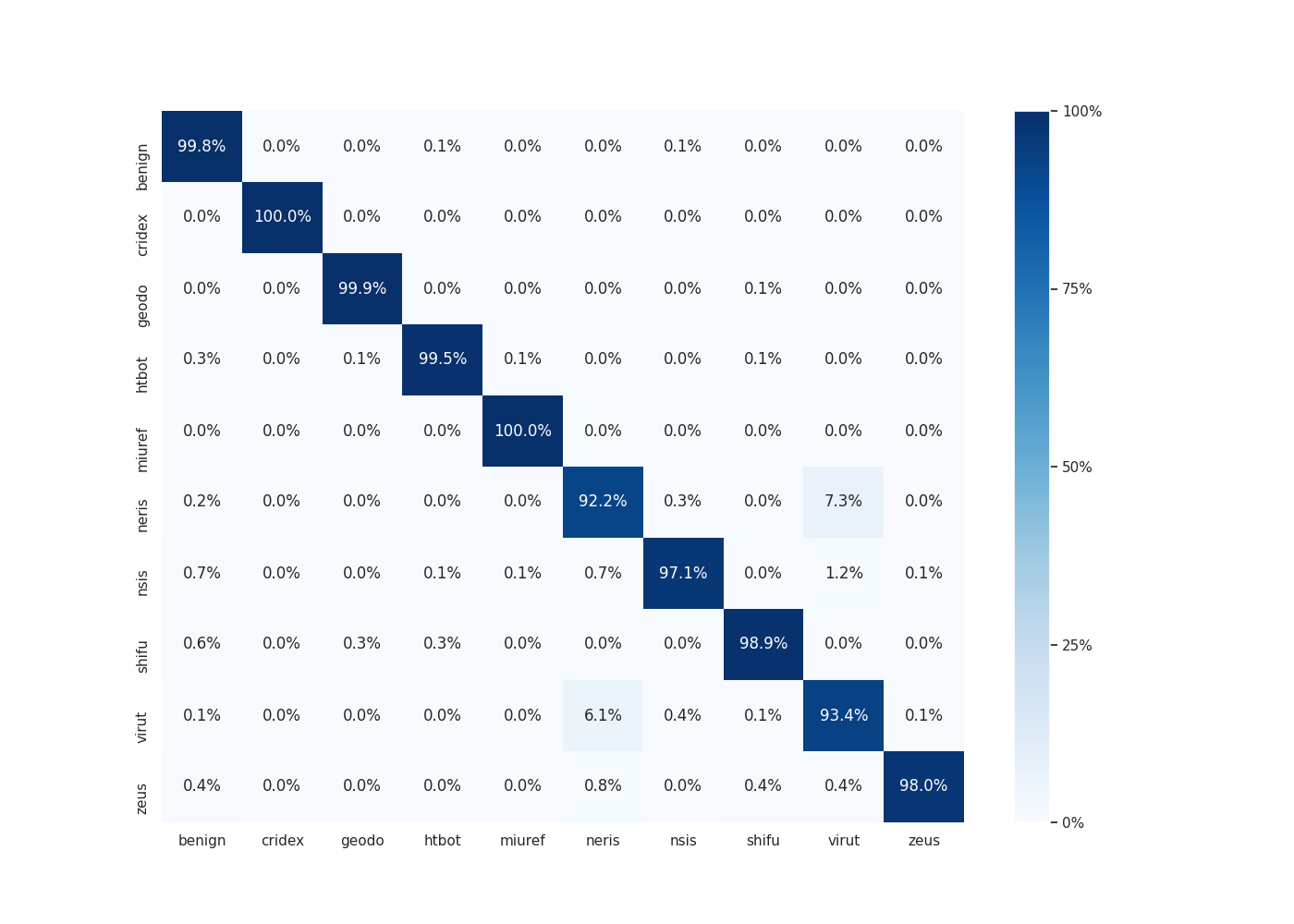}
}
\hfill
\subfloat[\textit{USTCB Random Forest}.\label{fig:cf-matrices-ustcb-rf}]{%
  \includegraphics[width=.49\textwidth]{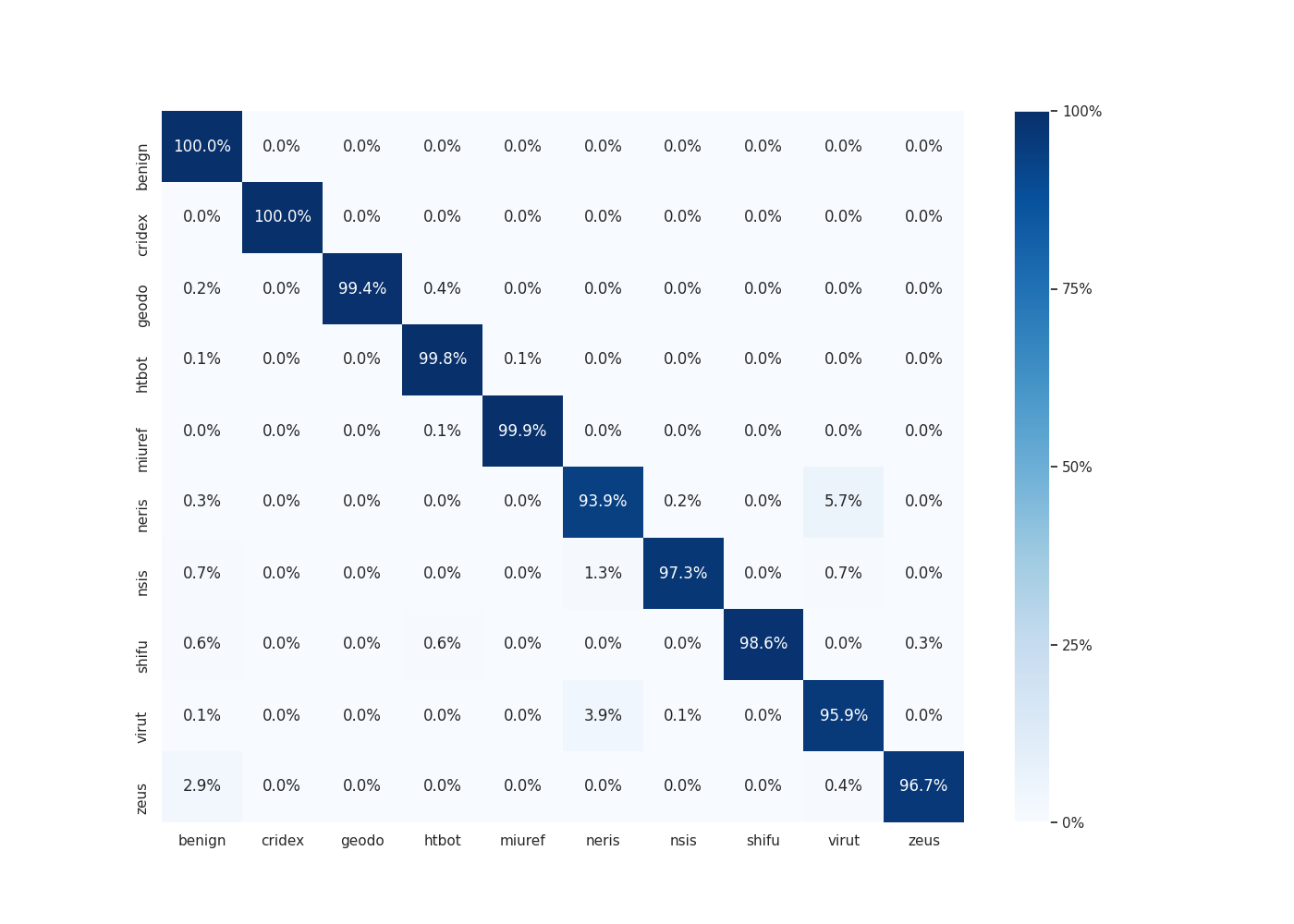}
}
\hfill
\subfloat[\textit{MUB MalDIST}.\label{fig:cf-matrices-mub-maldist}]{%
  \includegraphics[width=.49\textwidth]{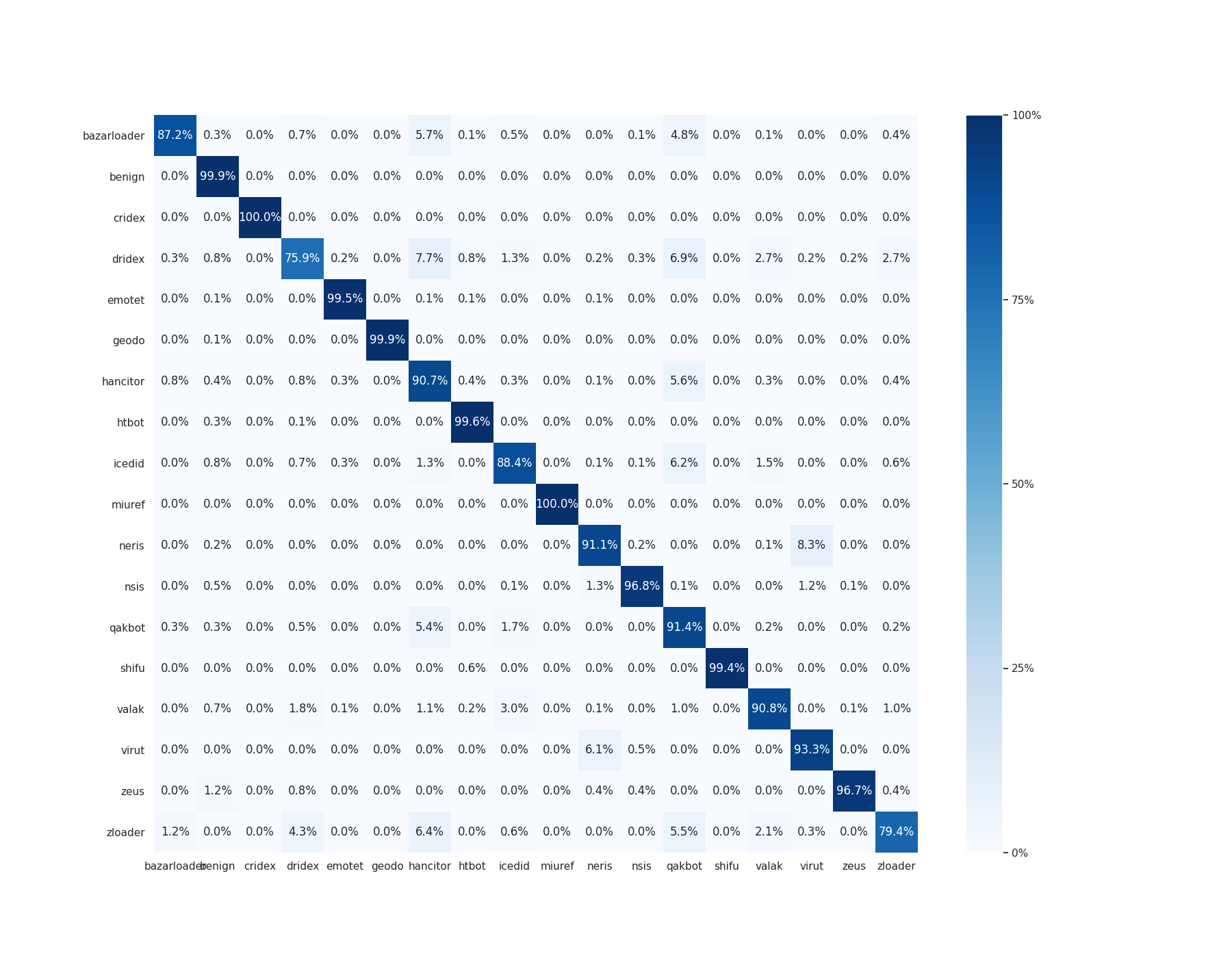}
  }
\hfill
\subfloat[\textit{MUB Random Forest}.\label{fig:cf-matrices-mub-rf}]{%
  \includegraphics[width=.49\textwidth]{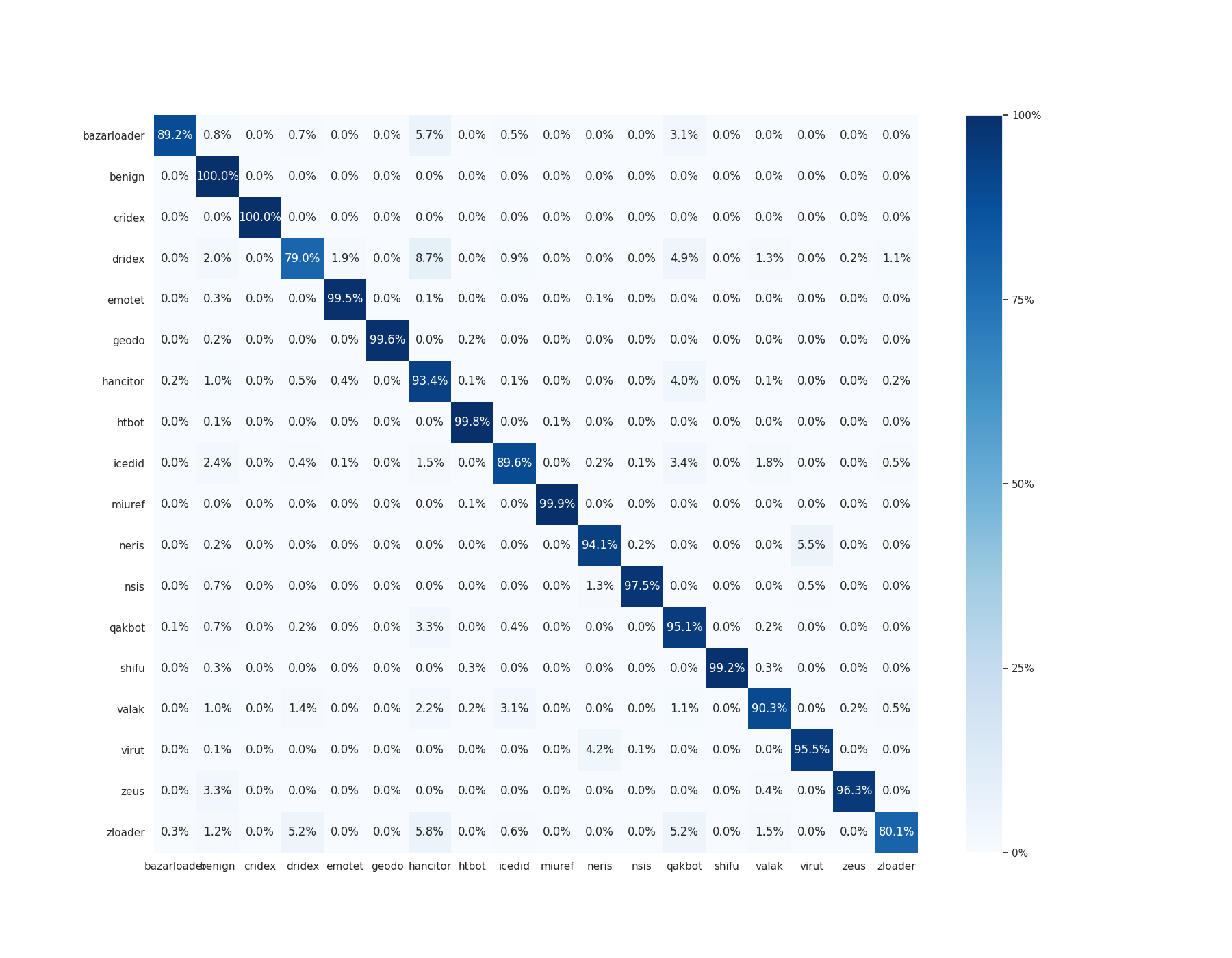}
  }
\caption{Confusion matrices of the two best models for the malware family traffic classification task. The rows are the true labels and the columns are the predicted labels.}
\label{fig:cf-matrices}
\end{figure*}

\subsection{Evaluation criteria}
For the evaluation of the models' performance (both ML and DL), we considered various metrics: Accuracy, Precision, Recall, and F1-score (F1), which are commonly used in the literature.  
Note that in this paper we tackle both classification problems on balanced datasets (i.e., malicious/benign) and imbalanced datasets (i.e., malware family). While accuracy is a useful metric for the first problem, it is not suitable for the latter. Thus, in the task of malware traffic family classification, we will also look at the precision, recall, and F1-score.

The complete definitions of the metrics are as follows:
\textbf{Accuracy}: the fraction of the total number of classification samples correctly classified.
    \[
        Accuracy = \frac{TP + TN}{TP + TN + FP + FN}
    \]
\textbf{Recall}: the total number of True Positives (TP) among all actual positive samples (TP + FN).
    \[
        Recall = \frac{TP}{TP + FN}
    \]
\textbf{Precision}: The ratio of True Positive (TP) samples of the total classifications that were positive (TP + FP).
    \[
        Precision = \frac{TP}{TP + FP}
    \]
\textbf{F1-score}:  a measure of a model’s accuracy on a dataset. It evaluates the binary classification systems, which classify samples as  positive or negative. The F1-score combines the Precision and Recall into one metric for the model's performance according to the harmonic mean of the model’s Precision and Recall.
    \[
        F1-score= 2 \cdot \frac{Precision \cdot Recall}{Precision + Recall}
    \]

    \begin{table}[hbtp]
 \caption{Detection (binary) of malware traffic classification in the following models: \textit{DeepMAL}, \textit{MalDIST}, \textit{M1CNN}, \textit{M2CNN}, \textit{RF}, \textit{KNN} and \textit{DT}}
\resizebox{\columnwidth}{!}{%
\begin{tabular}{l|ll|ll|ll|}
\cline{2-7}
                                       & \multicolumn{2}{c|}{\textbf{MTAB}}                       & \multicolumn{2}{c|}{\textbf{USTCB}}                      & \multicolumn{2}{c|}{\textbf{MUB}}                        \\ \cline{2-7} 
                                       & \multicolumn{1}{l|}{\textbf{Benign}}  & \textbf{Malware} & \multicolumn{1}{l|}{\textbf{Benign}}  & \textbf{Malware} & \multicolumn{1}{l|}{\textbf{Benign}}  & \textbf{Malware} \\ \hline
\multicolumn{1}{|l|}{\textbf{RF}}      & \multicolumn{1}{l|}{\textbf{100\%}}   & \textbf{99.60\%} & \multicolumn{1}{l|}{\textbf{99.90\%}} & \textbf{99.90\%} & \multicolumn{1}{l|}{\textbf{99.90\%}} & \textbf{99.90\%} \\ \hline
\multicolumn{1}{|l|}{\textbf{KNN}}     & \multicolumn{1}{l|}{96.80\%}          & 96.50\%          & \multicolumn{1}{l|}{98.50\%}          & 97.70\%          & \multicolumn{1}{l|}{98.20\%}          & 97.60\%          \\ \hline
\multicolumn{1}{|l|}{\textbf{DT}}      & \multicolumn{1}{l|}{99.40\%}          & 99.40\%          & \multicolumn{1}{l|}{99.70\%}          & 99.70\%          & \multicolumn{1}{l|}{99.50\%}          & 99.60\%          \\ \hline\hline
\multicolumn{1}{|l|}{\textbf{DeepMal}} & \multicolumn{1}{l|}{98.10\%}          & 98.20\%          & \multicolumn{1}{l|}{98.80\%}          & 99.10\%          & \multicolumn{1}{l|}{98.50\%}          & 98.90\%          \\ \hline
\multicolumn{1}{|l|}{\textbf{MalDIST}} & \multicolumn{1}{l|}{\textbf{99.90\%}} & \textbf{99.70\%} & \multicolumn{1}{l|}{\textbf{99.80\%}} & \textbf{99.90\%} & \multicolumn{1}{l|}{\textbf{99.90\%}} & \textbf{99.80\%} \\ \hline
\multicolumn{1}{|l|}{\textbf{M1CNN}}   & \multicolumn{1}{l|}{99.00\%}          & 98.50\%          & \multicolumn{1}{l|}{99.30\%}          & 99.20\%          & \multicolumn{1}{l|}{99.30\%}          & 99.20\%          \\ \hline
\multicolumn{1}{|l|}{\textbf{M2CNN}}   & \multicolumn{1}{l|}{98.70\%}          & 97.90\%          & \multicolumn{1}{l|}{98.00\%}          & 98.90\%          & \multicolumn{1}{l|}{98.80\%}          & 98.50\%          \\ \hline
\end{tabular}}
  \label{fig:binary}
\end{table}

\section{Experiments \& Results}
\label{sec:ER}
In the evaluation and comparison of the models, we implemented a five-fold cross-validation\footnote{Note that the cross-validation was on the data level, such that different models were evaluated on the exact same folds.} to obtain accurate and robust evaluations of the models with the three datasets, in the results we show the average of the folds. Our evaluation was based on both ML and DL models. The ML models were decision tree (DT), random forest (RF), and KNN (K=3), where KNN was used as a distance-based model while DT and RT are used as the tree-based models (where RF is an ensemble of decision trees). In terms of the DL models, we used DeepMAL, MalDIST, M1CNN\footnote{Note that, the M1CNN tackles the encrypted traffic classification, however, we were able to transfer the solution to the malware traffic classification domain with the same features and model.}, M2CNN, and MalDIST. The experiments were run on an Intel(R) Xeon(R) CPU E5-2683 v4 @ 2.10GHz with 64 GB RAM with 2 Geforce RTX 2080 Ti GPU\footnote{For the convenience of the reader, all the results also can be found in \textit{https://github.com/ArielCyber/When\_a\_RF\_Beats\_a\_CNN\_and\_GRU}}. 

\subsection{Malware Detection - Binary}
We first evaluated the performance of the models to differentiate between malware and benign instances. As half of each dataset consists of benign instances, and the other half malicious ones, the metric used for this evaluation was accuracy. In addition, we verified that in each fold of the cross-validation the portion of benign to malicious instances in the training and test sets were identical. Table~\ref{fig:binary} presents the results of binary malware traffic classification. The table, across all three datasets, demonstrates that KNN resulted in the worst accuracy while RF and MalDIST were characterized by the best accuracy (with a maximal difference of 0.1\% in their accuracy, across all datasets and classes).

\begin{figure*}[!]
\centering
\subfloat[USTCB dataset\label{fig:graph-ustcb-zero-day}]{%
      \includegraphics[width=2.8in]{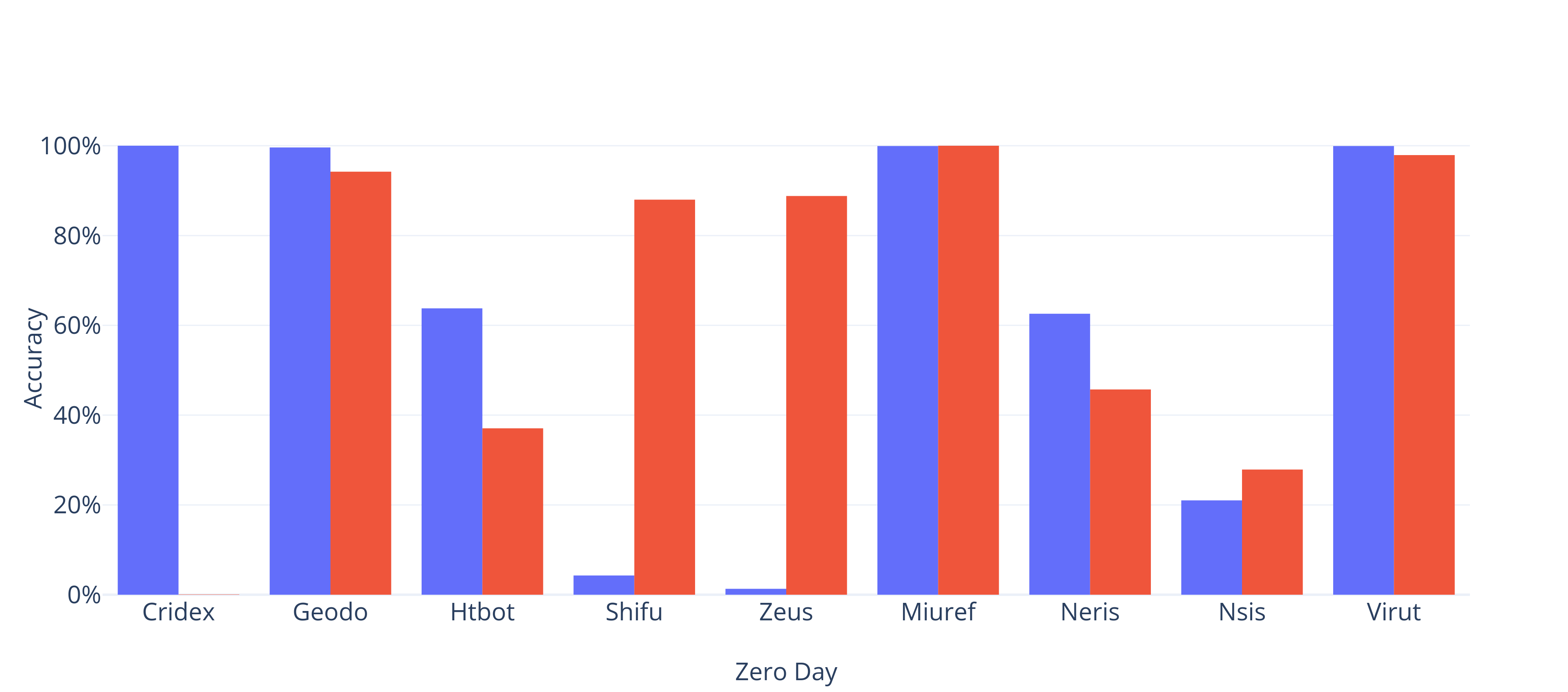}
    }
\subfloat[MTAB dataset\label{fig:graph-mtab-zero-day}]{%
      \includegraphics[width=2.8in]{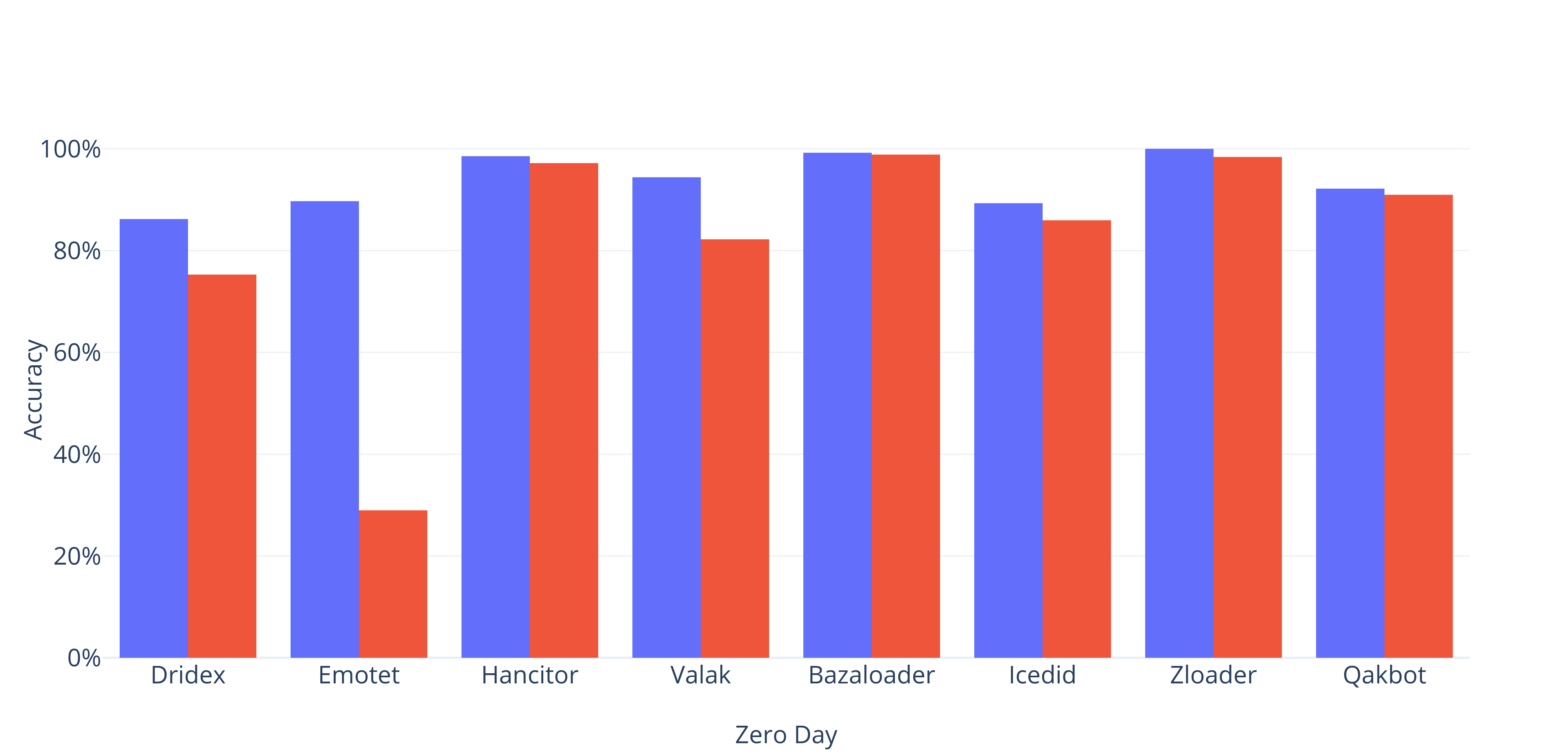}
    }
    
\subfloat[MUB dataset\label{fig:graph-mub-zero-day}]{%
      \includegraphics[width=2.8in]{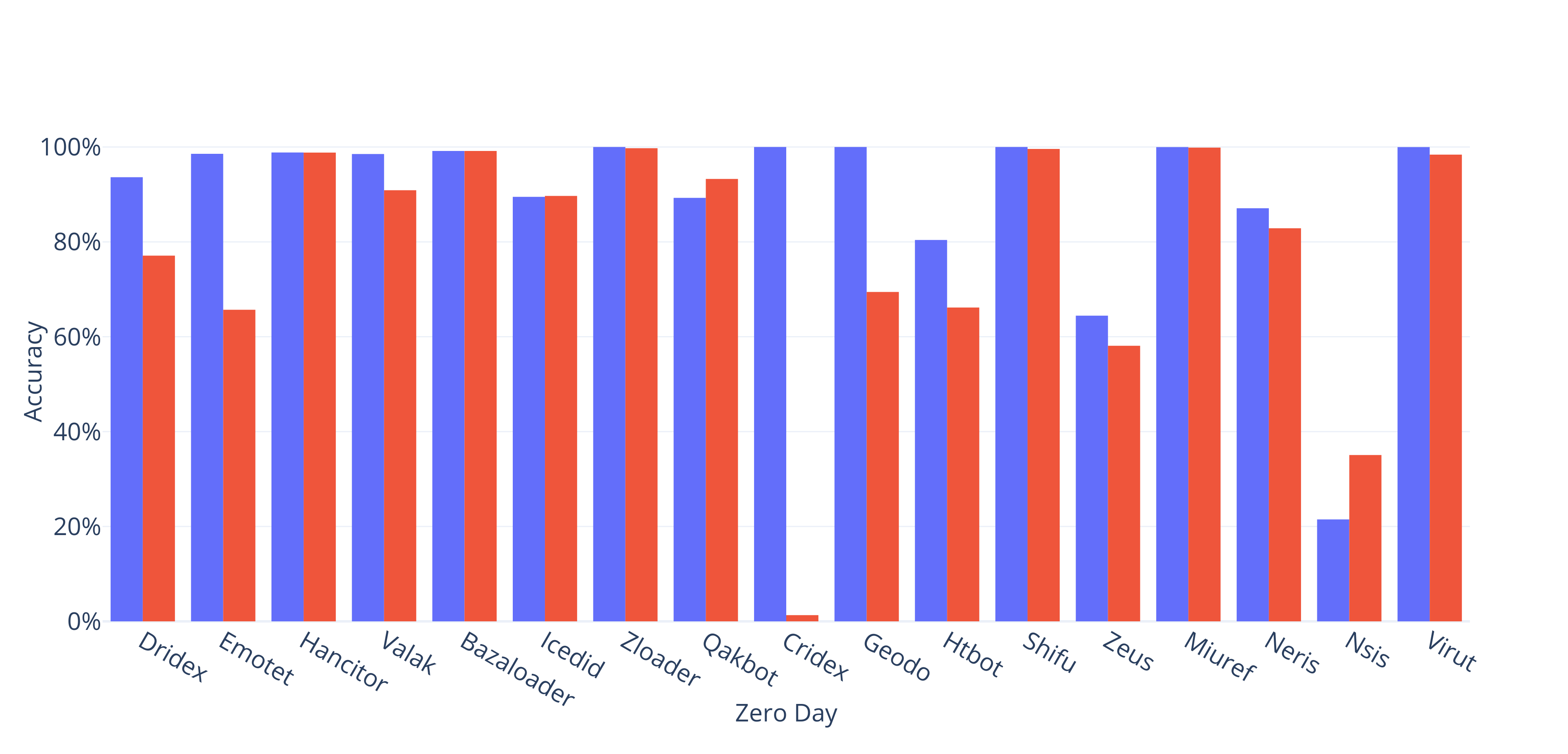}
    }

\includegraphics[width=0.2\textwidth]{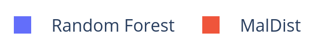}
\caption{Zero-day family results, the name of the tested family is presented on the x-axis}
\end{figure*}

\subsection{Malware Classification - Families}
Next, we extended our evaluation and analyzed whether the models can efficiently classify the malware families. Consequently, this evaluation only concerns the malicious part of each dataset. Since each class (family) has a different number of samples, in this evaluation we expanded the set of evaluation metrics to include precision, recall, and the F1-score as well. Figure~\ref{fig:malware-family-scores} depicts the results of this evaluation. As illustrated, this evaluation results in patterns similar to the benign-malicious one. In this case as well, the KNN models were shown to be the worst model, while RF and MalDIST were the best models in terms of classic ML and DL models, respectively. Therefore, we decided to focus only on the best models when analyzing their confusion matrices (Figures~\ref{fig:cf-matrices-mtab-maldist} -~\ref{fig:cf-matrices-mub-rf}). The analysis of the confusion matrices demonstrates that each model is better at classifying a different set of families. For example, for the confusion matrices on MTAB, MalDIST performed better than RF on Zloader, while RF performed better in classifying instances that belong to Dridex. We can also see that for the MUB dataset (i.e., the combination of the two sets of malwares) the same model had different detection rates of instances of the same family. For example, Shifu scores changed from 98.9\% in USTCB to 99.4\% in MUB, while the performance on Zloader dropped for MalDIST. The following question emerged as the result of this evaluation: Should DL-based models be used for malware family classification. Note that our results show an interesting phenomenon, whereby RF (i.e., the best classic model) dominated MalDIST (the best DL-based model) across all metrics and datasets, for the combined dataset (i.e., MUB) and even the basic decision tree models performed better than the best DL-based model. Note that all possible models cannot be evaluated in order to make a concrete statement. Furthermore, these findings do not result in the recommendation to avoid DL-based models, but rather exemplify that in some cases, there are more simplistic solutions, that may provide even better results than the DL-based models.

This brings us to question whether we truly need a complex DL architecture for this task? Perhaps RF is more than enough. To answer this query we continued the evaluation with our datasets and the two best models, RF and MalDIST.

\begin{figure*}[!]

\centering
\subfloat[Binary classification - MTAB dataset \label{fig:graph-mta-increase-malware-acc-label}]{%
\includegraphics[width=3.1in]{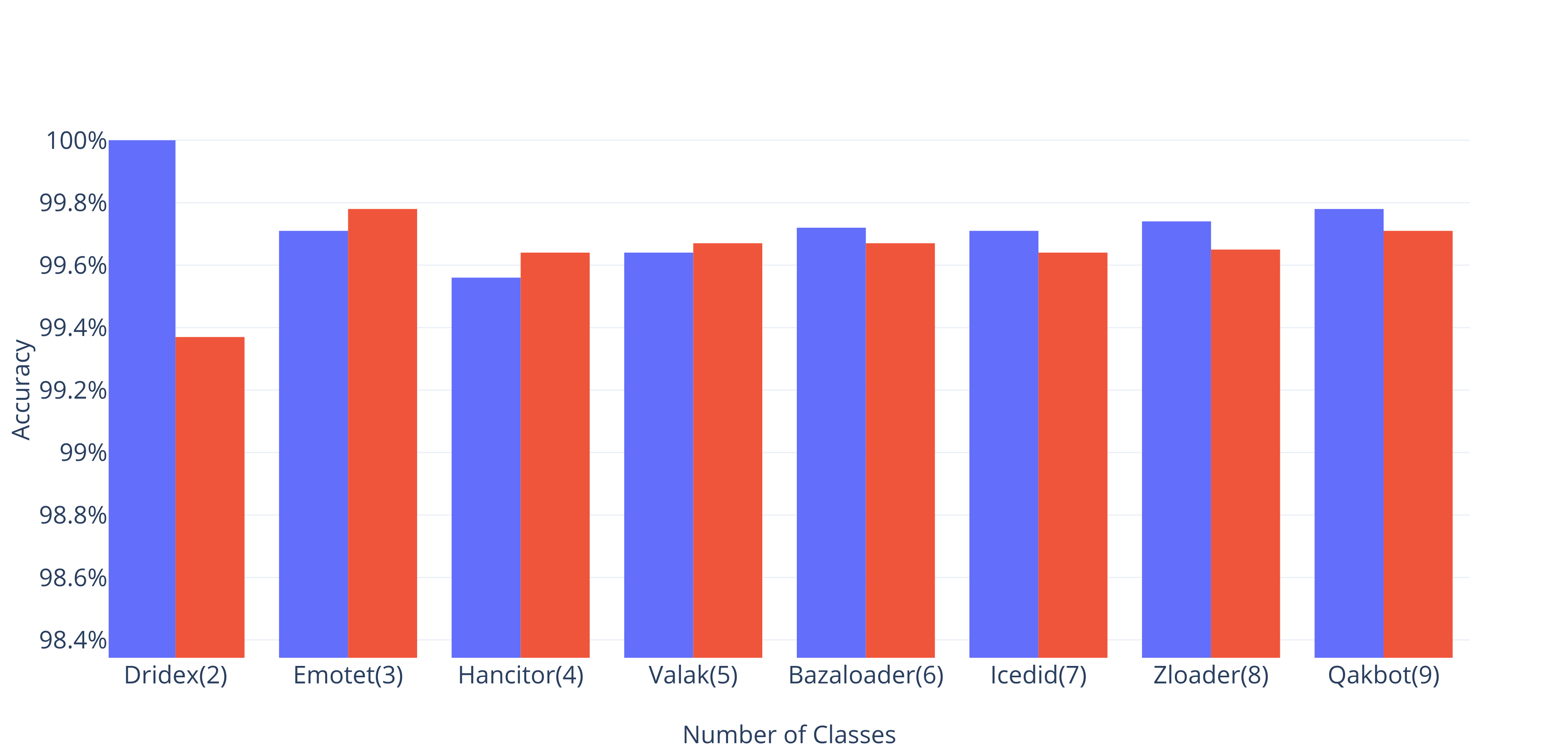}
    }
\subfloat[Multi-class family classification - MTAB dataset \label{fig:graph-mta-increase-malware-acc-malware-family}]{%
\includegraphics[width=3.3in]{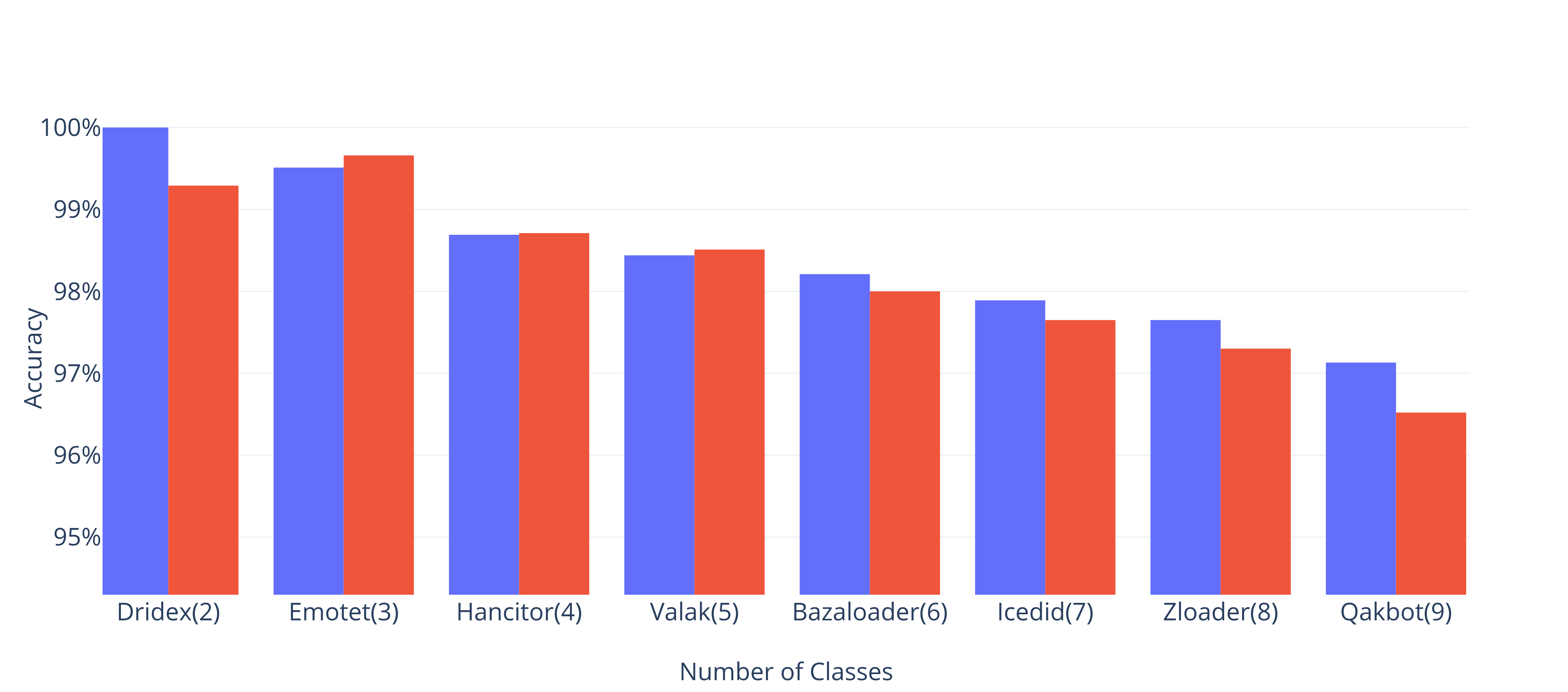}
    }
\hfill

\centering
\subfloat[Binary classification - USTCB dataset \label{fig:graph-ustc-increase-malware-acc-label}]{%
      \includegraphics[width=3.3in]{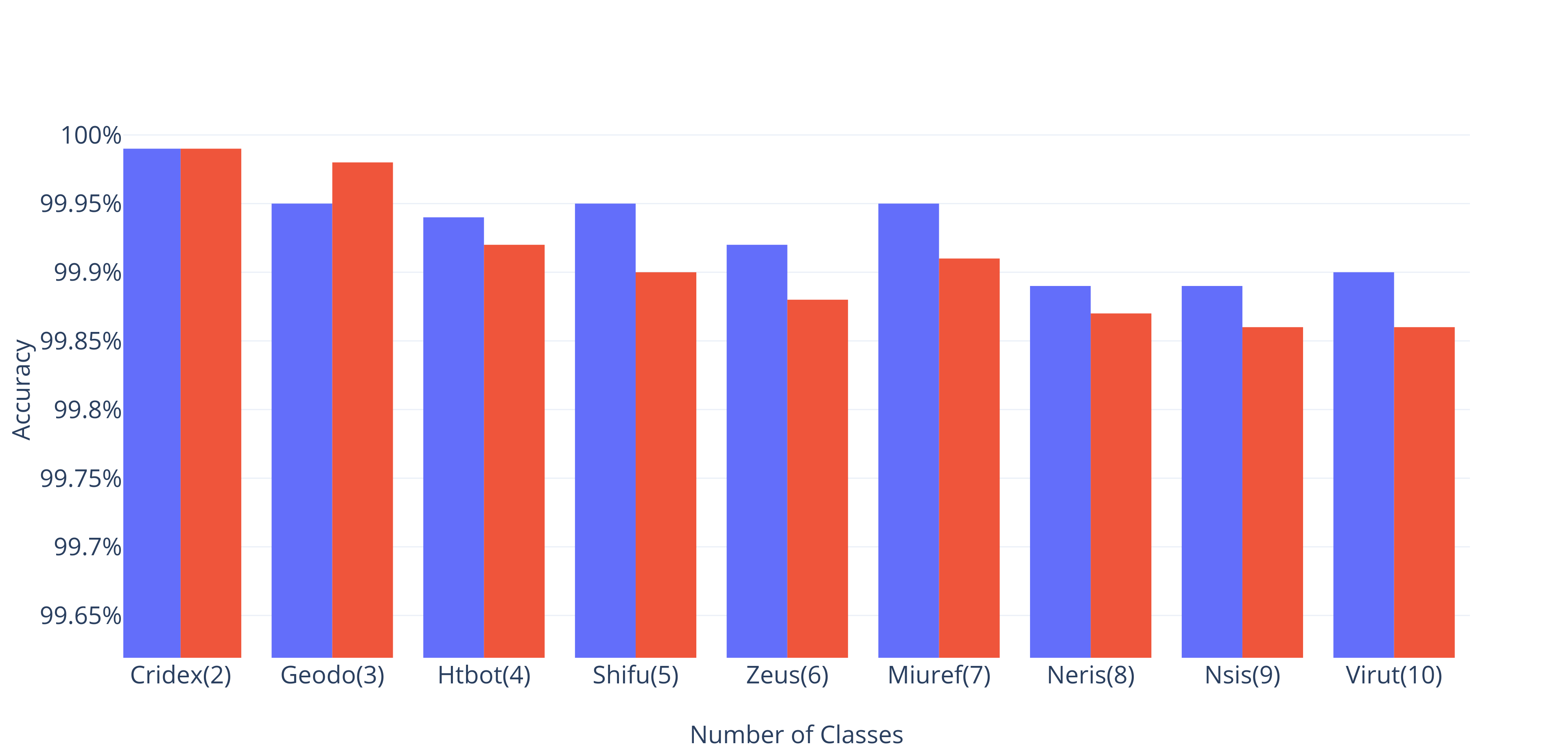}
    }
\centering
\subfloat[Multi-class family classification - USTCB dataset \label{fig:graph-ustc-increase-malware-acc-malware-family}]{%
      \includegraphics[width=3.3in]{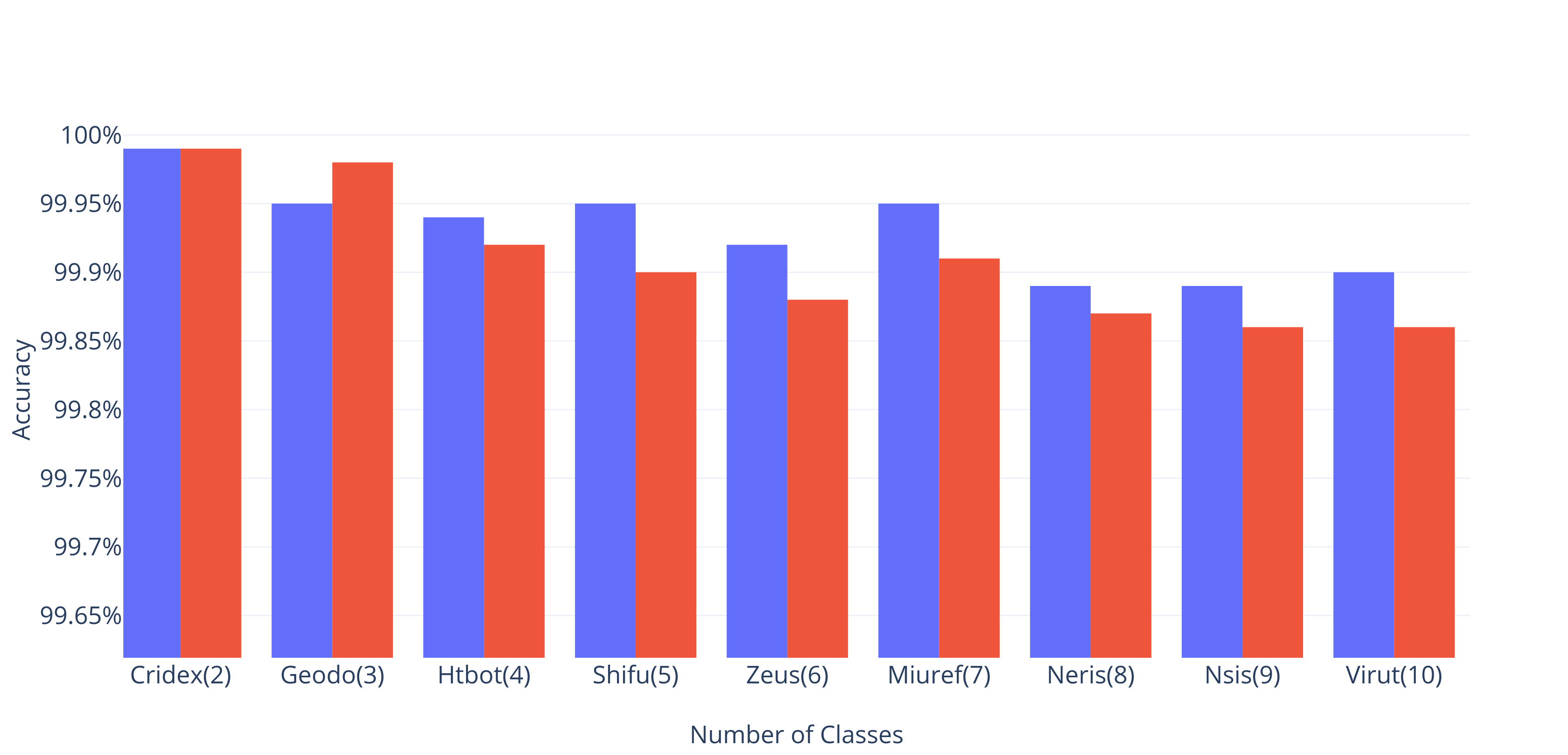}
    }
\hfill

\centering
\subfloat[Binary classification - MUB dataset \label{fig:graph-mub-increase-malware-acc-label}]{%
      \includegraphics[width=3.3in]{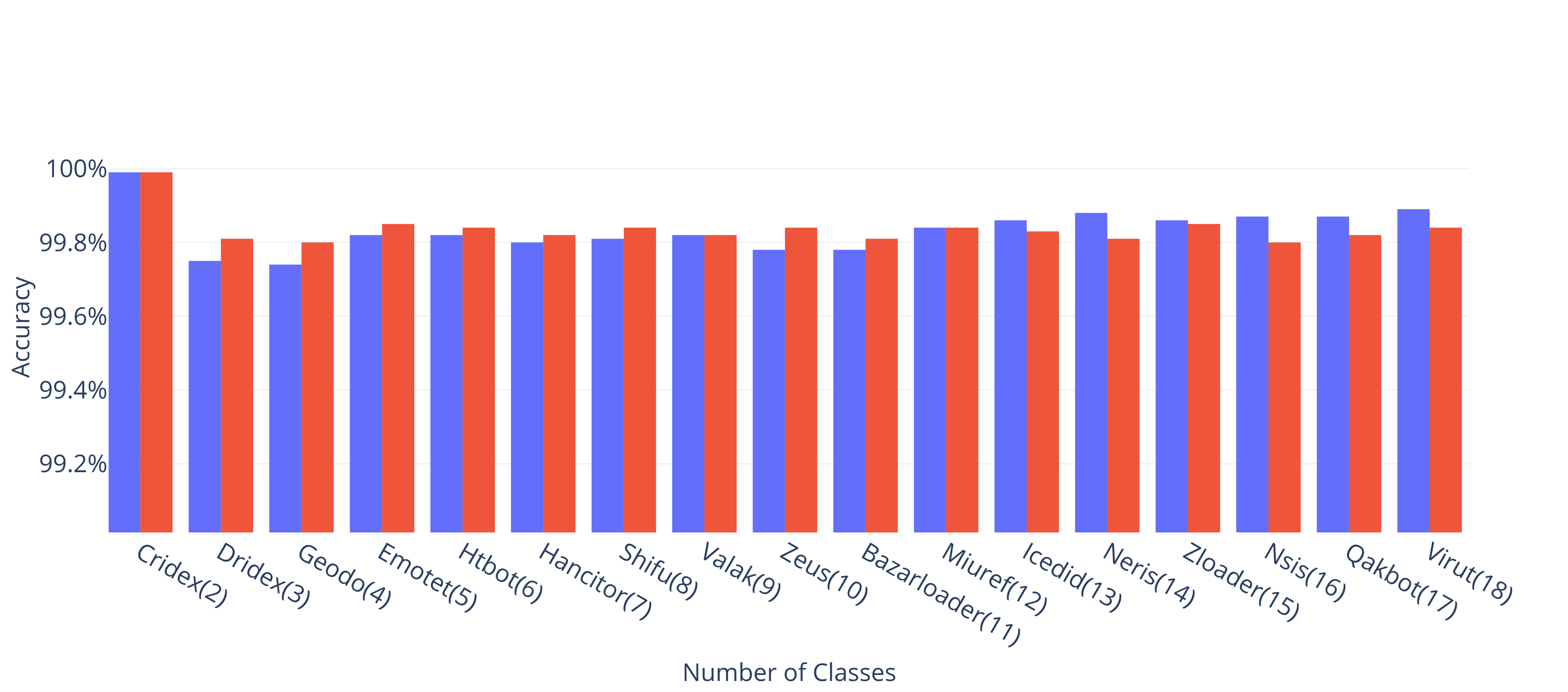}
    }
\centering
\subfloat[Multi-class family classification - MUB dataset \label{fig:graph-mub-increase-malware-acc-malware-family}]{%
      \includegraphics[width=3.3in]{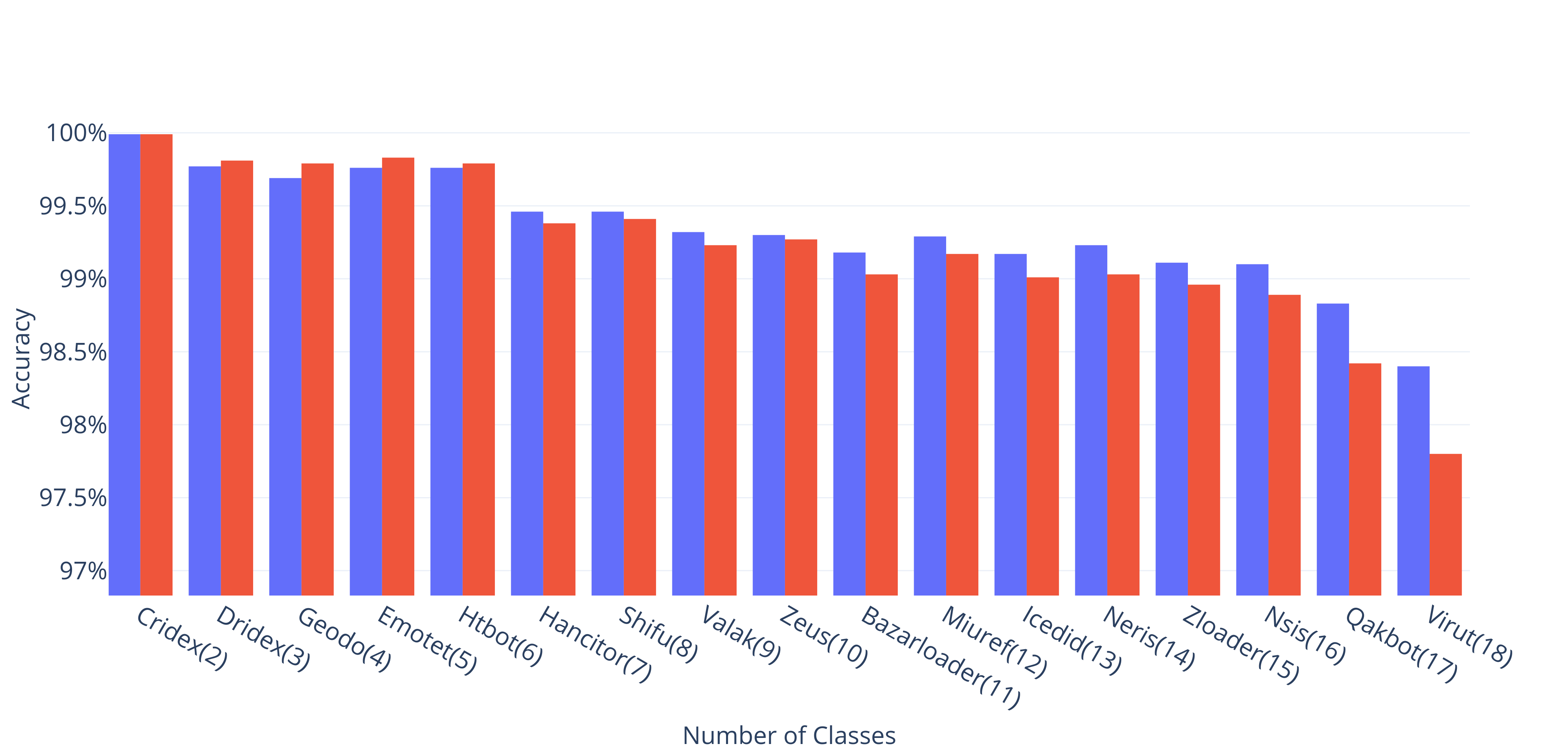}
    }
\hfill
  \centering
  \includegraphics[width=1.5in]{figures/Zero_Day_Legend.PNG}
  \caption{The influence of adding malware families to the dataset}
  \label{fig:graph-increase-malware-acc}
\end{figure*}

\subsection{Zero-day Classification}
Malware families keep changing on a daily basis. New families continuously appear, thus it is difficult to classify these instances, whereby a sufficient number of samples are collected to retrain the classifier with the updated set of samples. Thus, in the meantime, it is important that the model will be capable of at least classifying these new instances as malicious.
Our zero-day test is a binary classification test, where each time a different family is considered the test set, while the other families (including the benign instances) build the train set. In order to determine whether or not the model was able to identify the zero-day malware family, we focused on the accuracy score as our evaluation metric. Figure~\ref{fig:graph-mtab-zero-day} depicts the accuracy of RF and MalDIST on the detection of each family as a zero-day. Note that RF provided better results than MalDIST across all families, proving that in this case as well, a basic classic ML model is the best model for that task in this dataset.
On the other hand, when evaluating the same task on the USTCB dataset, the results were different. As depicted in Figure~\ref{fig:graph-ustcb-zero-day}, for some of the families (e.g., Geodo and Htbot), RF resulted in higher accuracy compared to MalDIST, while with other families (e.g., Nsis and Zeus), MalDIST performed better. An interesting insight is apparent in the extreme cases. For Cridex, RF resulted in about 90\% accuracy while MalDIST resulted in 0.6\%. On the other hand, for both Shifu and Zeus, while RF did not manage to classify these families instances as malicious, MalDIST did so quite easily. Figure~{\ref{fig:graph-mub-zero-day}} shows the same phenomenon for the the MUB dataset. For some of the families (e.g, Dridex, Emotet) RF provided better results than MalDIST and for other families such as Icedid, Qakbot, Nsis, the opposite was revealed. 

\subsection{Increase Malware}
Our final evaluation tackled the gradual increase of classes in the classified dataset. For each dataset, we iteratively added families and extracted the accuracy score on the resulted model. For example, in the $i$th step, the dataset consisted of instances of $i$ different families. The order of families in each dataset was randomly chosen. The MTAB started with Dridex, followed by Emotet, Hancitor, Valak, Bazarloader, Icedid, Zloader, and Qakbot. The USTCB started with Cridex, followed by Geodo, Htbot, Shifu, Zeus, Miuref, Neris, Nsis, and Virut. For the MUB we selected the families starting with Cridex, followed by Dridex, Geodo and ending with Virut\footnote{Note that the order in which the families are added is a combinatorial problem, which is not in the scope of this paper. We recommend investigating the effect of the different orders on the performance of the models in future research.}. 

First, as illustrated in Figures~\ref{fig:graph-mta-increase-malware-acc-label},~\ref{fig:graph-ustc-increase-malware-acc-label}, and~\ref{fig:graph-mub-increase-malware-acc-label} the performance of both models was steady and high, regardless of the number of classes added (i.e., the step $i$). Second, as depicted in Figures~\ref{fig:graph-mta-increase-malware-acc-malware-family},~\ref{fig:graph-ustc-increase-malware-acc-malware-family}, and~\ref{fig:graph-mub-increase-malware-acc-malware-family} for the family classification task a different pattern is apparent. In this task, we aimed to specifically classify each instance to its family and not just its maliciousness. Thus, as the number of families increased, the model had more options from which to choose and the task became harder. This is illustrated in the figures as the performance generally decreased as the number of families increased. The drop is more significant in the MTAB dataset compared to the USTCB one. Nonetheless, for both datasets, we notice that RF and MalDIST performed quite the same.

\begin{table}[H]
\caption{RF vs MalDIST}
\label{tab:our-datasets}
\begin{center}
\resizebox{\columnwidth}{!}{%
\begin{tabular}{ | m{1cm} | c | c | c | c |}
\hline
Datasets & model & normal acc & zero-day avg & increase malware avg \\
\hline
\multirow{2}{*}{MTAB} & MalDIST & 96.66\% & 82.24\% & 94.88\% \\ \cline{2-5}
& RF & 97.13\% & 93.7\% & 95.71\% \\ 
\hline
\multirow{2}{*}{USTCB} & MalDIST & 98.67\% & 64.41\% & 99.48\% \\ \cline{2-5}
& RF & 99.05\% & 61.39\% & 99.54\% \\ 
\hline
\multirow{2}{*}{MUB} & MalDIST & 98.40\% & 77.95\% & 99.22\% \\ \cline{2-5}
& RF & 97.79\% & 89.47\% & 99.34\% \\ 
\hline
\end{tabular}}
\end{center}
\end{table}

\section{Discussion and Conclusions}
\label{sec:con}
Internet traffic classification is an important task in terms of QoS, QoE, network visibility, and traffic-trend classification. While in the past the go-to solution for this task was to adapt machine learning, it has become somewhat obvious, that this solution should be based on deep learning architectures. Simple ML models have become to appear less and less in the literature and were thought of as inferior. In this paper, we ask the question, is this claim holds in practice, or should we enhance our tool-belt, and include classic ML models for classification tasks in this domain.
This paper validates our hypothesis that simple is not necessarily worse as we summarized these results in Table \ref{tab:our-datasets}. That is, adopting a complex DL architecture for the task of malicious traffic classification, does not guarantee the best performance. On the contrary, in some cases, classical machine learning algorithms such as DT or RL were more than enough. We validated our finding using two well-known malicious traffic datasets and a wide range of classification tasks (both binary and multi-class). First, for binary malware detection. While malicious families appear quite often, we show that even for the task of zero-day attack detection, the classic ML model does not fall short. 
Note that this paper focuses on the cybersecurity side of traffic classification. We recommend performing a  similar analysis for other tasks involving traffic classification (e.g., QoS and QoE). There is a question as to whether the same results will be attained for browser, operating system, or even application classification. At the end of the day, achieving more using more simplistic and less consuming tools is the end goal of the entire scientific community.

\section*{Acknowledgment}This work was supported by the Ariel Cyber Innovation Center in conjunction with the Israel National Cyber Directorate in the Prime Minister's Office.

\bibliographystyle{IEEEtran}
\bibliography{ref}

\end{document}